\documentclass[12pt]{amsart}
\usepackage{amsmath}
\usepackage{amsthm}
\usepackage{amsfonts}
\usepackage{amssymb}
\usepackage{epic}
\usepackage{eepic}

\theoremstyle{definition}

\theoremstyle{remark}

\newcommand{\beq}{\begin{equation}}
\newcommand{\eeq}{\end{equation}}

\newcommand{\pa}{\partial}
\newcommand{\ot}{\otimes}
\newcommand{\ra}{\rightarrow}

\newcommand{\ti}{\times}

\newcommand{\fr}[2]{{\textstyle \frac{#1}{#2} }}

\newcommand{\fsl}{{\mathfrak s}{\mathfrak l}}
\newcommand{\hfsl}{\widehat{\fsl}}

\newcommand{\bra}{\langle}
\newcommand{\ket}{\rangle}

\newcommand{\be}{\beta}
\newcommand{\ga}{\gamma}
\newcommand{\bga}{\bar{\gamma}}
\newcommand{\Ga}{\Gamma}
\newcommand{\de}{\delta}
\newcommand{\De}{\Delta}
\newcommand{\ep}{\epsilon}
\newcommand{\ka}{\kappa}
\newcommand{\la}{\lambda}
\newcommand{\om}{\omega}
\newcommand{\Om}{\Omega}
\newcommand{\si}{\sigma}
\newcommand{\up}{\Upsilon}
\newcommand{\vf}{\varphi}
\newcommand{\ve}{\varepsilon}

\newcommand{\bJ}{\bar{J}}

\newcommand{\bL}{\bar{L}}

\newcommand{\bv}{\Bar{v}}
\newcommand{\bw}{\bar{w}}
\newcommand{\bx}{\bar{x}}

\newcommand{\bz}{\bar{z}}

\newcommand{\CC}{{\mathcal C}}
\newcommand{\CD}{{\mathcal D}}

\newcommand{\CF}{{\mathcal F}}
\newcommand{\CG}{{\mathcal G}}

\newcommand{\CI}{{\mathcal I}}
\newcommand{\CJ}{{\mathcal J}}

\newcommand{\CN}{{\mathcal N}}
\newcommand{\CO}{{\mathcal O}}  
\newcommand{\CP}{{\mathcal P}}  
\newcommand{\CQ}{{\mathcal Q}}  
\newcommand{\CR}{{\mathcal R}}
\newcommand{\CS}{{\mathcal S}}
\newcommand{\CT}{{\mathcal T}}
\newcommand{\CU}{{\mathcal U}}
\newcommand{\CV}{{\mathcal V}}

\newcommand{\gv}{{\mathfrak v}}

\newcommand{\FJ}{{\mathfrak J}}
\newcommand{\BFJ}{\bar{{\mathfrak J}}}

\newcommand{\FP}{{\mathfrak P}}
\newcommand{\FR}{{\mathfrak R}}
\newcommand{\FS}{{\mathfrak S}}
\newcommand{\FT}{{\mathfrak T}}
\newcommand{\hfg}{\hat{{\mathfrak g}}}

\newcommand{\BN}{{\mathbb N}}
\newcommand{\BR}{{\mathbb R}}

\newcommand{\BC}{{\mathbb C}}

\newcommand{\BZ}{{\mathbb Z}}

\renewcommand{\Re}{\text{Re}}

\DeclareMathOperator{\sgn}{sgn}

\DeclareMathOperator{\Id}{Id}
\DeclareMathOperator*{\Res}{Res}

\newcommand{\CVHH}{\CV_{\text{\tiny HH}}}

\newcommand{\rf}[1]{(\ref{#1})}

\newcommand{\aufz}
{\begin{list}{$\bullet$}{\topsep0cm \itemsep0cm \parsep0cm}}
\newcommand{\eaufz}{\end{list}}
\newcounter{num}
\newcommand{\remlst}{\begin{list}
{(\arabic{num})}{\usecounter{num}\topsep0cm \itemsep0cm \parsep0cm}}
\begin{document}
\thispagestyle{empty}
\hspace*{\fill} DIAS-STP-99-07\\[.5cm]
\title{Operator product expansion and factorization
in the $H_3^+$-WZNW Model}
\author{\sc J. Teschner}
\address{School for theoretical Physics, Dublin Institute for Advanced
Studies, 10 Burlington Road, Dublin 4, Ireland}
\email{teschner@stp.dias.ie}

\begin{abstract}
Precise descriptions are given for the operator product 
expansion of generic primary fields as well as the factorization
of four point functions as sum over intermediate states. 
The conjecture underlying the recent derivation of the
space-time current algebra for string theory on $ADS_3$ 
by Kutasov and Seiberg is thereby verified. The
roles of microscopic and macroscopic states are further clarified.
The present 
work provides the conformal field theory prerequisites for a future study of 
factorization of amplitudes for string theory on $ADS_3$ as
well as operator product expansion in the corresponding 
conformal field theory on the boundary. 
\end{abstract}

\maketitle

\section{Introduction}
The $H_3^+$-WZNW model can be used to study 
string theory on backgrounds that contain an $ADS_3$-part.
Interest in such backgrounds was renewed\footnote{Earlier investigations
had studied mainly the physical spectrum via the $SU(1,1)$-WZNW model.
See e.g. \cite{ES}}
 by Maldacena's conjectures \cite{M}, 
which predict that string theory on backgrounds with $ADS_3$ should be
equivalent to a two dimensional conformal field theory on the boundary
of $ADS_3$.
One point that makes the study of the $ADS_3$ case of Maldacena's conjectures
particularly interesting is that the exact solvability of the
$H_3^+$-WZNW model \cite{Ga}\cite{T1} makes it possible to study
the ADS-CFT correspondence beyond the supergravity approximation 
\cite{GKS}\cite{KS}. More precisely, it becomes possible to 
investigate the regime
where the string coupling is small, but where the string length may be
of the same order as the curvature radius of $ADS_3$, which is a genuinely
``stringy'' regime.

Important steps in this direction were taken in the already mentioned papers
\cite{GKS}\cite{KS} by showing (among others) that the spectrum generating
algebra for string theory on $ADS_3$ (which is identified with the chiral 
algebra of the corresponding CFT on the boundary) 
can be constructed by worldsheet methods.
This was first done by free field methods in \cite{GKS} which is good enough
for the identification of observables and their transformation properties, as
well as the study of sectors describing strings stretched along the
boundary (as discussed in \cite{KS}\cite{SW}). In general one needs
to consider the full interacting $H_3^+$-WZNW model as was done in \cite{KS},
where the spectrum generating algebra was constructed in terms of 
primary fields and currents of the $H_3^+$-WZNW model. This construction
was based on a conjecture (eqn. (2.35) of \cite{KS}) on the OPE of 
a certain $H_3^+$-WZNW primary field $\Phi_1$ with a general primary field.
One result of the present paper will be to prove this conjecture on the
basis of the results of \cite{T1}. The crucial point is that the
contributions to the OPE of two primary field with 
OPE coefficients that contain delta-functions
follow unambigously from the ``smooth'' coefficients
determined in \cite{T1} via analytic continuation.
 
On the other hand, the author would like to take the opportunity to
complete the results of \cite{T1} in view of future applications such as
to string theory on $ADS_3$. This includes choosing a new normalization
for the primary field which is more natural from the point of view of 
string theory on $ADS_3$ than the one used in \cite{T1}, and 
describing how the operator product expansion (OPE) of two 
general primary fields is fully and
unambigously determined from the results in \cite{T1}. 

One of the main objectives of the present work however is to present a 
discussion of factorization for 
the four point function in the $H_3^+$-WZNW model which
should provide all ingredients needed from the $H_3^+$-WZNW model
for a future study of factorization and OPE in the {\it spacetime} CFT.
This includes a clarification of the roles of macroscopic
and microscopic states and their correspondence to operators.
It is sometimes claimed that state-operator 
correspondence breaks 
down in the $H_3^+$-WZNW model. This turns out to be rather misleading
in view of the results of the present paper. It is found that
the distinction between macroscopic and macroscopic states is in 
fact inessential as far as state-operator correspondence is concerned.
This will basically be a consequence of the remarkable analyticity
properties of the three point function found in \cite{T1}. 
The difference between microscopic and macroscopic states 
{\it is} of course essential for their appearance as intermediate
states in correlation functions. 

The paper is organized as follows: 
Section 2 summarizes results from \cite{T1}, rewritten using the new
normalization of primary fields. The motivation for choosing this 
normalization will be explained.
Besides that, the main new point of this section is a more
precise discussion of macroscopic and microscopic states.

Section 3 describes how to obtain the OPE in the general case from 
the special case presented in Section 2 via analytic continuation.
It is found that the OPE generically contains contributions which
involve delta-functions in the OPE coefficient. The conjecture
used in \cite{KS} is a particular case of the OPE's studied.

The fourth section then discusses factorization of the four point function.
It takes the form of an integral w.r.t. the label $j$ of the intermediate
representation, where the integrand factorizes into structure 
constants, two-point function and conformal blocks. 

The author has chosen to defer a couple of more technical aspects
to appendices in order to facilitate access to the main results and
ideas in the body of the paper. 

Appendix A provides some details on how to fix the normalization
and the relation between the normalizations used here and in \cite{T1}.

Appendix B demonstrates how the contributions to OPE and factorization
of four point function are determined by the current algebra symmetry.
The infinite-dimensionality of the zero mode representation introduces
some new features as compared to other CFT. The results of this 
appendix, although rather technical in nature, are of fundamental 
importance for the bootstrap approach to the $H_3^+$-WZNW model.

Appendix C finally discusses the Knizhnik-Zamolodchikov (KZ)
equations that correlation function of the $H_3^+$-WZNW model satisfy.

{\bf Acknowledgements} 
The author whishes to thank D. Kutasov and N.Seiberg for 
encouragement to publish his observation concerning 
the derivation of the OPE used in \cite{KS} from the results of \cite{T1},
as well as interesting remarks concerning string theory on $ADS_3$. 

\section{Some results on the $H_3^+$-WZNW model}

The $H_3^+$-WZNW is a conformal 
field theory that may be described by the Lagrangian
\begin{equation}
L=k(\pa\phi\bar{\pa}\phi+e^{2\phi}\bar{\pa}\ga\pa\bar{\ga}),
\end{equation}
see \cite{Ga} for more details on the Lagrangian formulation of this model.
The aim of the present section is to briefly summarize the picture that has
emerged from \cite{Ga}\cite{T1}\cite{T2}. Details 
on the relation of the normalizations used here and in \cite{T1} are given
in Appendix A.

\subsection{Canonical quantization}
Canonical quantization is performed by introducing canonical momenta 
$(\Pi,\be,\bar{\be})$ for the fields $(\phi,\ga,\bga)$ and defining the
operator algebra by canonical commutation relations. The space of states $\CV$
is then defined as tensor product of a Schr\"{o}dinger representation for
the zero modes $(\phi_0,\ga_0,\bga_0)$ on $L^2(H_3^+,e^{2\phi_0}
d\phi_0d^2\ga_0)$ with a representation of the non-zero modes on a Fock-space
$\CF$, see \cite{T1} for more details.\footnote{In comparison to 
\cite{T1} it should be noted that the fields $(\phi,\ga,\bga)$ and
$(\Pi,\be,\bar{\be})$ used here correspond in \cite{T1} to
$(-b\vf,bv,b\bv)$ and $(b\Pi_{\vf},b\Pi_{v},\Pi_{\bv})$ 
respectively, where
$b^{-2}=k-2$. Moreover, the central charge $k$ used here is the negative
of the central charge that appears there.}

States are represented by wave-functions $\Psi(h)\equiv
\Psi(\phi_0,\ga_0,\bga_0)$ taking values in $\CF$. There is a 
nondegenerate hermitian form on $\CF$ that makes the Hamiltonian symmetric,
but which is not positive definite.

The space of states $\CV$ carries a representation of two (left/right)
commuting isomorphic $\fsl_2$ current algebras generated by modes $J_n^a$ and
$\bJ_n^a$, $a=+,0,-$. The modes
$J_n^a$ satisfy the following commutation relations 
\begin{equation}\label{algebra}
\begin{aligned} {[}J_n^0,J_m^0{]}& =-\fr{k}{2}n\de_{n+m,0} \\ 
{[}J_n^0,J_m^{\pm}{]}& =\pm J_{n+m}^{\pm} \end{aligned} \qquad
{[}J_n^{-},J_m^{+}{]}=2J_{n+m}^0+kn\de_{n+m,0}.  
\end{equation}
 The expressions for the generators in terms of
the canonical fields contains terms proportional to $e^{-2\phi_0}$ which
reflect the interaction of the $H_3^+$-WZNW model. These terms disappear
for $\phi_0\ra\infty$, in which case one recovers free field representations
of the $\fsl_2$ current algebra. These free field representations differ
from the usual free field representations by a different treatment of the
zero modes: The zero mode representation space is $\CS(\BC)$ and the
generators $\ga_0,\bga_0$ act as multiplication operators, their 
conjugate momenta $\be_0,\bar{\be}_0$ as the corresponding derivatives.
 
As usual one has associated with the 
current algebras two commuting Virasoro algebras with generators 
$L_n$, $\bL_n$ by means of the Sugawara construction:
\begin{equation}\label{Sugawara}\begin{split}
L_m=&\frac{1}{2(k-2)}\sum_{k\in\BZ}:-2J^0_{-k}J^0_{m+k}+J^{+}_{-k}J^{-}_{m+k}
+J^{-}_{-k}J^{+}_{m+k}:\\
& \text{where}\qquad 
:J_{m}^aJ_n^b:=\begin{cases}J_{m}^aJ_n^b & \text{if $m<n$}\\
\frac{1}{2}(J_n^aJ_n^b+J_n^bJ_n^a) & \text{if $m=n$}\\
J_{n}^bJ_m^a & \text{if $m>n$}, 
\end{cases}
\end{split}
\end{equation}
and correspondingly for the $\bL_n$. 

\subsection{Spectrum}
The decomposition of $\CV$ into irreducible representations \cite{Ga}
can be written as 
\begin{equation}\label{absdec}
\CV\simeq\int\limits_{\CC^+}^{\oplus} dj \;\,\CR_{j} ,
\end{equation}
where $\CC^+=-\frac{1}{2}+i\BR^+$ and the representations $\CR_{j}$ 
are defined as follows:
One starts with a representation $P_j$ for the zero modes $J_0^a$,
$\bJ_0^a$ which corresponds to a principal series representation of 
$SL(2,\BC)$. It may be realized e.g. on
the Schwartz-space $\CS(\BC)$ of functions on $\BC$ 
by means of the differential
operators $\CD_j^a$
\begin{equation}
\CD^+_j=-x^2\pa_{x}+2jx \qquad \CD^0_j=-x\pa_{x}+j 
\qquad \CD^-_j=-\pa_{x},
\end{equation}
together with their complex conjugates $\bar{\CD}_j^a$ on $\CS(\BC)$.
This representation of the zero mode algebra is then extended to a 
representation of the full current algebra by 
requiring $J_n^a P_j=0=\bJ_n^a P_j$, $n>0$, and generating 
the space $\CR_{j}$ by acting with the generators $J_n^a$, $\bJ_n^a$, $n<0$.

\subsection{Microscopic vs. macroscopic states}
\subsubsection{Macroscopic states}
It is expected that a convenient plane wave basis for $\CV$
can be constructed in terms of wave functions $\Psi(j;x|h)$, 
$j\in -\frac{1}{2}+i\BR^+$, $x\in\BC$, that are
uniquely characterized by the asymptotic behavior
\begin{equation}\label{asym}
\Psi\bigl(-\fr{1}{2}+i\rho;x|h\bigr) \sim e^{-\phi_0}\Bigl(
e^{-2i\rho\phi_0}\de^2(\ga_0-x)+B\bigl(-\fr{1}{2}+i\rho\bigr)e^{2i\rho\phi_0}
|\ga_0-x|^{4i\rho-2} \Bigr)\Om,
\end{equation}
together with their descendants obtained by acting with generators $J_{-n}^a$, 
$\bJ_{-n}^a$. The function $B(j)$ that appears in \rf{asym} will 
be given later in \rf{Bcoeff}.

The corresponding ``states'' $|j;x\ket$ and their
duals $\bra j;x|$ are more precisely 
to be understood as distributions on dense subspaces $\FS\subset \CV$.
From the asymptotic behavior \rf{asym} one would expect them to be
delta-function normalizable, so following the terminology of \cite{S}
one may call them ``macroscopic states''. It indeed turns out 
(cf. the Appendix A) that the normalization implied by \rf{asym} is 
equivalent to 
\begin{equation}\label{normstates}
\bra -\fr{1}{2}-i\rho_2;x_2|-\fr{1}{2}-i\rho_1;x_1\ket
=
\de^{(2)}(x_2-x_1)\de(\rho_2-\rho_1).
\end{equation}

\subsubsection{Microscopic states}
The wave-functions $\Psi(j;x|h)$ are expected to possess an analytic
continuation w.r.t. the variable $j$ \footnote{See \cite{T3} for some
discussion of the corresponding issue in Liouville theory}. The 
``states'' $|j;x\ket$, $\bra j;x|$ for $j\notin -\fr{1}{2}+i\BR$
are not even delta-function normalizable (cf. \rf{asym}) and will be called
``microscopic states''. The ``microscopic states'' can still 
be understood as distributions on dense subspaces $\FT_j\subset \CV$,
but their domains $\FT_j$ will be considerably smaller than those of the
macroscopic states, e.g. not include subspaces $\FS\subset \CV$
like the Schwartz-space. 
The difference between microscopic states and macroscopic states 
is crucial for their possible appearance in the spectrum: The
non-appearance of microscopic states in spectral decompositions follows 
from rather general arguments as e.g. summarized in \cite{Be}. However, this
distinction will 
turn out (cf. Section 3.3.2 below) {\it not to be relevant} 
for the possibility of 
having a suitable generalization of state-operator correspondence in
the $H_3^+$-WZNW model.

\subsubsection{Description of microscopic states in a spectral 
representation} 
In a spectral representation \rf{absdec} one may characterize $\bra j,x|$ as
follows. It is defined on the subspace $\FT_j\subset\CV$ of vectors
\begin{equation}\label{psiexp}
|\Psi\ket=\int\limits_{\CC^+}
dj'\int\limits_{\BC}d^2x \; 
\Psi_0(j';x)|j';x\ket+\text{descendants}
\end{equation}
where $\Psi_0(j';x)$ is analytic in a connected domain that contains the axis $
-\frac{1}{2}+i\BR^+$ and has the point $j'=j$ in its closure, and such that
$\lim_{j'\ra j}\Psi_0(j';x)\equiv \Psi_0(j;x)$ exists. The value of the 
linear form $\bra j,x|$ on $|\Psi\ket$ is then given by $\bra j,x|\Psi\ket=
\Psi_0(j;x)$. 

\subsubsection{Generalization} 
More generally one may define $\CVHH$ to be
the linear space consisting of vectors of the form
\begin{equation}\label{micrdec}\begin{aligned}
|\Psi\ket= & |\Psi\ket_{\text{\tiny macr}}+|\Psi\ket_{\text{\tiny micr}} \\
=& \int\limits_{\CC^+}
dj\int\limits_{\BC}d^2x \; 
\Psi_0(j;x)|j;x\ket
+\sum_{j\in\CI}\int\limits_{\BC}d^2x \Psi_0(j;x)
|j;x\ket +\text{descendants},
\end{aligned}\end{equation}
where $\CI$ is some {\it finite} subset of $\BC\setminus -\frac{1}{2}+i\BR$.
The notation $\CVHH$ (HH for Hartle-Hawking) is used in analogy to 
the notation used in \cite{S} for the space of microscopic states
in Liouville theory.

An ``inner product'' of two microscopic states $\bra \Psi_2|$ and 
$|\Psi_1\ket$ can then be defined whenever 
\begin{enumerate}
\item the sets $\CI_2$ and $\CI_1$ defined by writing $\bra \Psi_2|$ and 
$|\Psi_1\ket$ as in \rf{micrdec} are disjoint,
\item ${}_{\text{\tiny macr}}
\bra \Psi_2|$ is in the domain of $|\Psi_1\ket_{\text{\tiny micr}}$ and 
$|\Psi_1\ket_{\text{\tiny macr}}$ is in the domain of 
${}_{\text{\tiny micr}}\bra \Psi_2|$, 
and finally
\item ${}_{\text{\tiny macr}}
\bra \Psi_2|\Psi_1\ket_{\text{\tiny macr}}$ exists. 
\end{enumerate}
It is then given by
\begin{equation}
\bra \Psi_2|\Psi_1\ket={}_{\text{\tiny macr}}
\bra \Psi_2|\Psi_1\ket_{\text{\tiny micr}}+
{}_{\text{\tiny micr}}
\bra \Psi_2|\Psi_1\ket_{\text{\tiny macr}}+
 {}_{\text{\tiny macr}}
\bra \Psi_2|\Psi_1\ket_{\text{\tiny macr}}
\end{equation}
Although these definitions certainly call for further generalization
and refinement, they will turn out to be sufficient for 
a consistent interpretation of microscopic intermediate states in 
correlation functions.

\subsection{Primary fields}

\subsubsection{}
One is interested in operators $\Phi^j(x|z)$, $x,z\in\BC$ 
having the following characteristic OPE
\begin{equation}\label{prim2}
  J^a(z)\Phi^j(x|w)=\frac{1}{z-w}      \CD_j^a \Phi^j(x|w),\qquad 
\bJ^a(\bz)\Phi^j(x|w)=\frac{1}{\bz-\bw}\bar{\CD}_j^a \Phi^j(x|w).
\end{equation}
The operators $\Phi^j(x|z)$ are primary also w.r.t. the Sugawara Virasoro 
algebra with conformal dimensions 
\begin{equation}
\De_j=-\frac{1}{k-2}j(j+1).
\end{equation}
\subsubsection{}
Such operators can semiclassically be identified 
with the following functions of the 
variables that appear in the $H_3^+$-Lagrangian:
\begin{equation}\label{classfield}
\Phi^j(x|z)=\frac{2j+1}{\pi}\Bigl((\ga-x)(\bar{\ga}-\bar{x})e^{\phi}+e^{-\phi}
\Bigr)^{2j}.
\end{equation}
Normal ordering will not allow the quantum operators to have 
such a simple form. They should however simplify for $\phi\ra\infty$
where the interaction vanishes:
\begin{equation}\label{opasym}
\Phi^j(x|z) \;\sim \;\;:e^{2(-j-1)\phi(z)}:\de^2(\ga(z)-x)+B(j):e^{2j\phi(z)}:
|\ga(z)-x|^{4j}.
\end{equation}
The operator $\de^2(\ga-x)$ makes sense when smeared over $x$ since functions
of $\ga(z)$ do not require normal ordering. 

\subsubsection{}
One would expect that such operators can be constructed for any {\it complex}
value of $j$. Operator-state correspondence 
\begin{equation}\label{stopprim}
\lim_{z\ra 0}\Phi^{j}(x|z)|0\ket = |j;x\ket \qquad
\lim_{z\ra\infty}|z|^{4\De_j}\bra 0|\Phi^{-j-1}(x|z)=\bra j;x|
\end{equation} 
is then to be understood in the distributional sense, i.e. when 
evaluated against states in $\FT_j$. 

\subsubsection{}
The asymptotic expression \rf{opasym} fixes a normalization
for $\Phi^j$ which will be motivated in the following subsection.
It will be shown in Appendix A that this normalization
is equivalent to the two point function
\begin{equation}\label{twopt}\begin{aligned}
\bra \Phi&^{-\frac{1}{2}-i\rho'} (x_2|z_2) 
\Phi^{-\frac{1}{2}+i\rho}(x_1|z_1)\ket
=\\
=& |z_2-z_1|^{-4\De_{\rho}}\Bigl(
\de^{(2)}(x_2-x_1)\de(\rho'-\rho)+B\bigl(-\fr{1}{2}+i\rho\bigr)
|x_2-x_1|^{2(2i\rho-1)}\de(\rho'+\rho)\Bigr),
\end{aligned}
\end{equation}
where the coefficient function $B(j)$ is explicitely given as 
\begin{equation}\label{Bcoeff}
 B(j)= -\bigl(\nu(b)\bigr)^{2j+1}\frac{2j+1}{\pi}
\frac{\Ga(1+b^2(2j+1)\bigr)}{\Ga(1-b^2(2j+1)\bigr)}, 
\quad \nu(b)=\pi\frac{\Ga(1-b^2)}{\Ga(1+b^2)},\quad b^2=\frac{1}{k-2}.
\end{equation}
\subsubsection{}
There is a linear relation between the
operators $\Phi^j(x|w)$ and $\Phi^{-j-1}(x|w)$:
\begin{equation}\label{refrel}
\Phi^{j}(x|z)=R(j)(\CI_j\Phi^{-j-1})(x'|z),
\end{equation}
where 
the {\it reflection amplitude} $R(j)$ is given by
\begin{equation}\label{refamp}
R(j)=-\bigl(\nu(b)\bigr)^{2j+1}
\frac{\Ga(1+b^2(2j+1)\bigr)}{\Ga(1-b^2(2j+1)\bigr)},
\end{equation}
and $\CI_j$ is the intertwining operator that establishes
the equivalence of the $SL(2,\BC)$-representations 
$P_{-j-1}$ and $P_j$:
\begin{equation}
(\CI_j\Phi^{-j-1})(x|z)=\frac{2j+1}{\pi}\int_{\BC}d^2x'\; |x-x'|^{4j}
\Phi^{-j-1}(x'|z).
\end{equation}
The operator $\CI_j$ 
is normalized such that $\CI_{-j-1}\circ \CI_j=\Id$, which implies 
its unitarity for $j\in -\frac{1}{2}+i\BR$. This means that the 
map between in- and out-states created by the operator $\Phi^j(x|z)$
according to \rf{stopprim} is also unitary for these values of $j$.  

\subsection{Normalization}
For applications to string theory on $ADS_3$ one is interested in negative 
values of $j$ since $-j\equiv h$ acquires the interpretation of being
the scaling dimension of an operator in the CFT living on the boundary.
As one hopes this CFT to be stable, one should only have positive $h$ 
in the spectrum. 

The vertex operators in the $H_3^+$-WZNW model can be used as ingredients
for string theory vertex operators that describe scattering of strings
created by sources on the boundary. For example, there is a class of
vertex operators that take the form
\begin{equation}\label{vertop}
V_n(x)=\int_{\BC}d^2z \;\Phi^{j}(x|z)\,\Xi_n(z); \qquad \De_j+\De_n=1,
\end{equation}
where $\Xi_n$ is a spinless worldsheet operator in the CFT describing $\CN$.
Such vertex operators are interpreted as describing a string state created
by a {\it pointlike} source located at the point $x$ of the boundary  
\cite{BORT}, \cite{KS}.

For this interpretation one needs that the leading behavior of the 
vertex operator $\Phi^{j}(x|z)$ for $\phi\ra \infty$ is proportional to
$\de^2(\ga-x)$, which is in fact found from the semiclassical expression
\rf{classfield} (cf. loc. cit.). It is then natural to normalize the vertex
operators $\Phi^{j}(x|z)$ to have asymptotics \rf{opasym}.
Semiclassically this corresponds to using the fields \rf{classfield}.

\subsection{Three point function}
The three point function of the vertex operators $\Phi^j(x|w)$ normalized
by \rf{twopt} takes the form
\begin{equation}\label{threept}\begin{aligned}
 \bra\Phi^{j_3}(x_3|z_3) & \Phi^{j_2}(x_2|z_2)
  \Phi^{j_1}(x_1|z_1)\ket= D(j_3,j_2,j_1)C(j_3,j_2,j_1|x_3,x_2,x_1)\cdot\\
 & \cdot
|z_1-z_2|^{2(\De_3-\De_2-\De_1)}|z_1-z_3|^{2(\De_2-\De_1-\De_3)}
|z_2-z_3|^{2(\De_1-\De_2-\De_3)}. 
\end{aligned}\end{equation}
\subsubsection{Structure constants $D(j_3,j_2,j_1)$}
\begin{equation}\label{Strconst}
D( j_3 ,j_2,j_1) =
\frac{G(j_1+j_2+j_3+1)G(j_1+j_2-j_3)G(j_1+j_3-j_2)
G(j_2+j_3-j_1)}
{(\nu(b))^{-j_1-j_2-j_3-1}\; G_0 \;G(2j_1+1)G(2j_2+1)G(2j_3+1)}.
\end{equation}
The special function $G(j)$ is related to the $\Upsilon$-function
introduced in \cite{ZZ} via $G(j)=b^{-b^2x(x+1+b^{-2})}\Upsilon^{-1}(-bj)$ 
and 
\begin{equation} \label{consts}
G_0=-2\pi^2\frac{\Ga(1+b^2)}{\Ga(-b^2)}G(-1)
\end{equation}
The $\Upsilon$-function was in loc. cit. defined
by an integral representation which converges in the strip $0<\Re(x)<Q/2$ 
and defines an analytic function
there. It may alternatively be constructed out of the Barnes 
Double Gamma function\cite{Ba}\cite{Sh} as follows:
\[
\begin{aligned}{} & \Upsilon^{-1} (s)= 
\Ga_2(s|b,b^{-1})\Ga_2(b+b^{-1}-s|b,b^{-1}),\\
&\log\Ga_2(s|\om_1,\om_2)=  \lim_{t\ra 0}\frac{\pa}{\pa t}
\sum_{n_1,n_2=0}^{\infty}
(s+n_1\om_1+n_2\om_2)^{-t}.
\end{aligned}
\]
\subsubsection{
The coefficients $C(j_3,j_2,j_1|x_3,x_2,x_1)$} 
\begin{equation}\label{OPEcoef}
C(j_3,j_2,j_1|x_3,x_2,x_1)= 
|x_1-x_2|^{2(j_1+j_2-j_3)}
|x_1-x_3|^{2(j_1+j_3-j_2)}
|x_2-x_3|^{2(j_2+j_3-j_1)}
\end{equation}
They are Clebsch-Gordan coefficients\footnote{More precisely:
distributional kernel} for the 
decomposition of the tensor product $P_{j_2}\ot P_{j_1}$ of 
$SL(2,\BC)$ principal 
series representations, cf. \cite{N}.  

The $x_i$-dependence of the $C(j_3,j_2,j_1|x_3,x_2,x_1)$ is 
uniquely
fixed by $SL(2,\BC)$-invariance as long as none of $j_1+j_2-j_3$,
$j_1+j_3-j_2$, $j_2+j_3-j_1$ equals $-n-1$, $n=0,1,2,\ldots$. 

\subsection{Operator product expansion}
The following result \cite{T1} provides the basis for a complete determination 
of the operator product expansion for any product 
$\Phi^{j_2}\Phi^{j_1}$,
$\Phi^{j_i}=\Phi^{j_i}(x_i,\bx_i|z_i,\bz_i)$, $i=1,2$:

There is a range $\CR$ for the values of $j_2$, $j_1$ given by the 
inequalities
\begin{equation}\label{funran}
|\Re(j_{21}^{\pm})|<\fr{1}{2}
\qquad\qquad j_{21}^+=j_2+j_1+1,\qquad j_{21}^{-}=j_2-j_1.
\end{equation}
such that the operator product expansion takes the form
\begin{equation}\label{OPE}\begin{aligned}
\Phi^{j_2}(x_2|z_2) \Phi^{j_1}(x_1|z_1)
= 
\int_{\CC^+}dj_3  & \;\,D_{21}(j_3)\;|z_2-z_1|^{\De_{21}(j_3)}
\;\bigl(\CJ_{21}(j_3)\Phi^{-j_3-1}\bigr)(z_1)\\
   & +\text{descendants},
\end{aligned}
\end{equation}
where the operator $(\CJ_{21}(j_3)\Phi^{-j_3-1})(z_1)$ has been defined as
\begin{equation}\label{defphi}
\bigl(\CJ_{21}(j_3)\Phi^{-j_3-1}\bigr)(z_1)
\equiv \int_{\BC}d^2x_3 \;\,C(j_3,j_2,j_1|x_3,x_2,x_1)
\Phi^{-j_3-1}(x_3|z_1)
\end{equation}
and the abbreviations $D_{21}(j_3)=D(j_3,j_2,j_1)$ and 
$\De_{21}(j_3)=\De_{j_3}-\De_{j_2}-\De_{j_1}$ have been introduced for
later convenience.
A more precise description of the descendant contributions is given
in Appendix B.

The region \rf{funran} will turn out to be 
precisely the maximal region in which
$j_1$, $j_2$ may vary such that none of the poles of 
the integrand in \rf{OPE} hits the contour $\CC$
of integration over $j_3$ in the OPE.

\section{Analytic continuation of the OPE}

The aim of the present section is to demonstrate that the operator 
product expansion \rf{OPE} admits an analytic continuation to 
generic complex values of $j_1$, $j_2$.
 
\subsection{Analytic properties of the integrand in \rf{OPE}}

\subsubsection{Reflection property}
It will be important to note that the integrand 
of the $j_3$-integration in \rf{OPE} is symmetric under $j_3\ra -j_3-1$.
This can be seen as a consequence of the reflection relation \rf{refrel}
when used to express the operator $\Phi^{-j_3-1}$ 
in terms of $\Phi^{j_3}$. The integral over $x_3$ in \rf{defphi} may then 
be carried out by using the integral (see e.g. \cite{Do})
\begin{equation}\label{Doint1}
\int_{\BC}d^2t \;\,|t|^{2a}|1-t|^{2b}=-\pi \frac{\ga(-1-a-b)}{\ga(-a)\ga(-b)},
\qquad \ga(x)=\frac{\Ga(x)}{\Ga(1-x)}.
\end{equation}
The resulting functional
equation  for $(\CJ_{21}(j_3)\Phi^{-j_3-1})(z_1)$ can be written as  
\begin{equation}\label{reflPhi}
 \begin{aligned}
\bigl(  \CJ_{21}(j_3)  &\Phi^{-j_3-1}\bigr)(z_1)=\\
& =-\frac{1}{\pi}
\frac{\ga(-2j_3)}{\ga(j_1-j_2-j_3)\ga(j_2-j_1-j_3)}
\frac{1}{B(j_3)}
\bigl(\CJ_{21}(-j_3-1)
\Phi^{j_3}\bigr)(z_1).
 \end{aligned}
\end{equation}
The structure constants on the other hand satisfy the identity 
\begin{equation}\label{refstruct}
D_{21}(j_3)=-\pi\frac{\ga(j_1-j_2-j_3)\ga(j_2-j_1-j_3)}{\ga(-2j_3)}
B(j_3)D_{21}(-j_3-1),
\end{equation}
which can be easily verified using the functional equations $
G(x)=G(-x-1-b^{-2})$ and
\begin{equation}\label{funcrel}
G(j-1)=\frac{\Ga(1+b^2j)}{\Ga(-b^2j)}G(j), \qquad
G(j-b^{-2})=b^{2(2j+1)}\frac{\Ga(1+j)}{\Ga(-j)}G(j),
\end{equation}
which follow from the corresponding functional equations of the $\up$-function
given in \cite{ZZ}. The relations \rf{reflPhi} and \rf{funcrel}
imply the symmetry of the integrand in \rf{OPE} under $j_3\ra -j_3-1$.

\subsubsection{Poles of the structure constants}
To obtain the meromorphic contination of the structure
constants $D(j_3,j_2,j_1)$ one may note that
the function $G(j)$ admits a meromorphic continuation to the complex
$j$ plane which may be defined by the functional relations \rf{funcrel}.
It follows that
the function $G(j)$ has poles for $j=n+mb^{-2}$ and 
$j=-(n+1)-(m+1)b^{-2}$, $n,m=0,1,2,\ldots$. The resulting 
set of poles of the structure constants is given by
\begin{equation}
\begin{aligned}
j_3= & j_{21}^{+}-1-n-mb^{-2} \\
j_3= & j_{21}^{-}-(n+1)-(m+1)b^{-2} \\
j_3= & -j_{21}^{\pm}-(n+1)-(m+1)b^{-2}
\end{aligned}
\quad
\begin{aligned}
j_3= & j_{21}^{+}+n+(m+1)b^{-2} \\
j_3= & j_{21}^{-}+n+mb^{-2}\\
j_3= & -j_{21}^{\pm}+n+mb^{-2}
\end{aligned}\quad n,m=0,1,2,\ldots
\end{equation}
 
\subsubsection{Poles of the operator $\CJ_{21}(j_3)\Phi^{-j_3-1}$}
Concerning the meromorphic continuation of the operator 
$(\CJ_{21}(j_3)\Phi^{-j_3-1})(z_1)$ one should first note that the 
distribution $|x|^{2j}$ is meromorphic in $j$ with
poles at $j=-n-1$, $n=0,1,2,\ldots$ and residues
\begin{equation}\label{res}
\Res_{j=-n-1}|x|^{2j}=\frac{\pi}{(n!)^2}\de^{(n,n)}(x)\equiv
\frac{\pi}{(n!)^2}(\pa_x\pa_{\bx})^n\de(x),
\end{equation}
see e.g. \cite{GS}. One consequently finds 
poles of the OPE coefficients for $j_3=j_{21}^{+}+n$, 
$j_3=\pm j_{21}^{-}-n-1$. 

A further series of poles arises from the 
integration over $x_3$ in \rf{defphi}. These are related to the poles
from the distributions $|x|^{2j}$ by the reflection $j_3\ra -j_3-1$.

\subsubsection{}
It is important to note that the descendant contributions 
to the OPE do not introduce
further poles with positions depending on $j_1$, $j_2$,
as follows from the results in Appendix B.

Alltogether one finds the following set $\FP(j_2,j_1)$ 
of poles ($n,m=0,1,2,\ldots$)
\begin{equation}\label{polestot}
\begin{aligned}
j_3= & j_{21}^{\pm}-1-n-mb^{-2} \\
j_3= & -j_{21}^{\pm}-1-n-mb^{-2}
\end{aligned}
\qquad\qquad 
\begin{aligned}
j_3= & j_{21}^{\pm}+n+mb^{-2} \\
j_3= & -j_{21}^{\pm}+n+mb^{-2}
\end{aligned}
\end{equation}
 It is useful to visualize the position of the poles in the complex
$j_3$-plane:

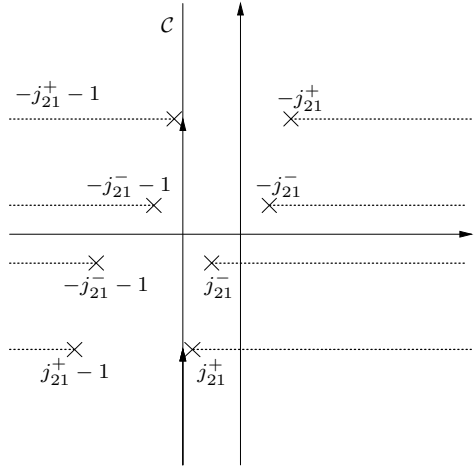
\begin{center}
\setlength{\unitlength}{0.0005in}
\begingroup\makeatletter\ifx\SetFigFont\undefined%
\gdef\SetFigFont#1#2#3#4#5{%
  \reset@font\fontsize{#1}{#2pt}%
  \fontfamily{#3}\fontseries{#4}\fontshape{#5}%
  \selectfont}%
\fi\endgroup%
{\renewcommand{\dashlinestretch}{30}
\begin{picture}(4824,4839)(0,-10)
\put(1962,912){\makebox(0,0)[lb]{\smash{{{\SetFigFont{8}{14.4}{\rmdefault}{\mddefault}{\updefault}$j_{21}^+$}}}}}
\path(1812,12)(1812,1212)
\blacken\path(1842.000,1092.000)(1812.000,1212.000)(1782.000,1092.000)(1842.000,1092.000)
\path(1812,12)(1812,1212)
\blacken\path(1842.000,1092.000)(1812.000,1212.000)(1782.000,1092.000)(1842.000,1092.000)
\path(1812,12)(1812,1212)
\blacken\path(1842.000,1092.000)(1812.000,1212.000)(1782.000,1092.000)(1842.000,1092.000)
\path(1812,1212)(1812,3612)
\blacken\path(1842.000,3492.000)(1812.000,3612.000)(1782.000,3492.000)(1842.000,3492.000)
\path(1812,3537)(1812,4812)
\path(2412,12)(2412,4812)
\blacken\path(2442.000,4692.000)(2412.000,4812.000)(2382.000,4692.000)(2442.000,4692.000)
\dottedline{45}(1512,2712)(12,2712)
\dottedline{45}(912,2112)(12,2112)
\dottedline{45}(2712,2712)(4737,2712)
\dottedline{45}(2112,2112)(4737,2112)
\path(1437,2637)(1587,2787)
\path(1437,2787)(1587,2637)
\path(2637,2637)(2787,2787)
\path(2787,2637)(2637,2787)
\path(837,2037)(987,2187)
\path(987,2037)(837,2187)
\path(2037,2037)(2187,2187)
\path(2037,2187)(2187,2037)
 \dottedline{45}(1727,3612)(12,3612)
\dottedline{45}(687,1212)(12,1212)
\dottedline{45}(2937,3612)(4812,3612)
\dottedline{45}(1917,1212)(4812,1212)
\path(2862,3537)(3012,3687)
\path(3012,3537)(2862,3687)
 \path(1647,3537)(1797,3687)
 \path(1797,3537)(1647,3687)
\path(612,1137)(762,1287)
\path(762,1137)(612,1287)
\path(1832,1137)(1992,1287)
\path(1832,1287)(1992,1137)
\put(1587,4512){\makebox(0,0)[lb]{\smash{{{\SetFigFont{8}{14.4}{\rmdefault}{\mddefault}{\itdefault}$\CC$}}}}}
\put(62,3802){\makebox(0,0)[lb]{\smash{{{\SetFigFont{8}{14.4}{\rmdefault}{\mddefault}{\updefault}$-j_{21}^+-1$}}}}}
\put(800,2862){\makebox(0,0)[lb]{\smash{{{\SetFigFont{8}{14.4}{\rmdefault}{\mddefault}{\updefault}$-j_{21}^--1$}}}}}
\put(562,1812){\makebox(0,0)[lb]{\smash{{{\SetFigFont{8}{14.4}{\rmdefault}{\mddefault}{\updefault}$-j_{21}^--1$}}}}}
\put(337,912){\makebox(0,0)[lb]{\smash{{{\SetFigFont{8}{14.4}{\rmdefault}{\mddefault}{\updefault}$j_{21}^+-1$}}}}}
\put(2787,3762){\makebox(0,0)[lb]{\smash{{{\SetFigFont{8}{14.4}{\rmdefault}{\mddefault}{\updefault}$-j_{21}^+$}}}}}
\put(2562,2862){\makebox(0,0)[lb]{\smash{{{\SetFigFont{8}{14.4}{\rmdefault}{\mddefault}{\updefault}$-j_{21}^-$}}}}}
\put(2037,1812){\makebox(0,0)[lb]{\smash{{{\SetFigFont{8}{14.4}{\rmdefault}{\mddefault}{\updefault}$j_{21}^-$}}}}}
\path(12,2412)(4812,2412)
\blacken\path(4692.000,2382.000)(4812.000,2412.000)(4692.000,2442.000)(4692.000,2382.000)
\end{picture}
}

\end{center}
The crosses symbolize the first poles to the left (right) of the contour
$\CC$, the dashed lines indicate the lines along which further poles follow
with spacing $1$ and $b^{-2}$.

\subsection{Continuing the integral over $j_3$ in \rf{OPE}}
When attempting to 
continue beyond the region given by \rf{funran} one finds poles that hit
the contour of integration $\CC^+$. However, as long as the 
imaginary parts of $j_{21}^{+}$ and $j_{21}^{-}$ do not vanish
one may define the analytic continuation of \rf{OPE} by deforming 
the contour $\CC^+$. The deformed contour may always be represented
as the sum of the original contour and finitely many
circles around the poles. These latter lead to a finite sum of 
residue contributions to the OPE.
The important case where $j_{21}^{\pm}$ are real can be
treated by giving them a small imaginary part which is sent to zero
after having deformed the contour. The result does 
not depend on the sign of the imaginary part
as a consequence of the symmetry 
of the integrand in \rf{OPE} under $j_3\ra -j_3-1$.

Only the case $j_{21}^{+}<-\frac{1}{2}$ will be described explicitly
here since
the generalization to the other cases is straightforward. 
For this case one finds an expression of the 
following form:
\begin{equation}\label{OPEfull}\begin{aligned}
\Phi^{j_2} (x_2|z_2) \Phi^{j_1}(x_1|z_1) 
=& \sum_{j_3\in \CP^+_{21}}
z_{21}^{\De_{21}(j_3)}\;
D_{21}^{}(j_3)\;\bigl(\CJ_{21}^R(j_3)\Phi^{-j_3-1}\bigr)
(z_1)  \\
+& \sum_{j_3\in \CP^-_{21}}\;
z_{21}^{\De_{21}(j_3)}
D^{R}_{21}(j_3)\;
\bigl(\CJ_{21}^{}(j_3)\Phi^{-j_3-1}\bigr)
(z_1)\\
+&\sum_{\ve\in\{+,-\}}\sum_{j_3\in \CQ^{\pm}_{21}}
z_{21}^{\De_{21}(j_3)}
D^{R}_{21}(j_3)
\bigl(\CJ_{21}^{}(j_3)\Phi^{-j_3-1}\bigr)
(z_1)\\
+ &
\int_{\CC^+}dj_3 \;\,z_{21}^{\De_{21}(j_3)}\;
D_{21}(j_3)\;\bigl(\CJ_{21}(j_3)\Phi^{-j_3-1}\bigr)
(z_1)\\
+ &\text{descendants},
\end{aligned}
\end{equation}
where the ranges of summations are given by the sets 
\begin{equation}\begin{aligned}
\CP^{\pm}_{21}=&\{ j_3=j_{21}^{\pm} +n\,|0\leq n; \Re(j_3)<-\fr{1}{2} \},\\
\quad \CQ^{\pm}_{21}=&\{ j_3=j_{21}^{\pm} +n+(m+1)b^{-2}\,|0\leq n,m;
\Re(j_3)<-\fr{1}{2} \},
\end{aligned}\end{equation}
and $D_{21}^R(j)$, $\CJ_{21}^R(j)$ denote the residues of 
$D_{21}(j)$, $\CJ_{21}(j)$ when $j$ is a pole of these functions.

It is interesting to note that the residue $\CJ_{21}^R(j_{21}^++n)$
contains delta-functions as a consequence of 
\rf{res}:
\begin{equation}
\CJ_{21}^R(j_{21}^++n)=\frac{\pi}{(n!)^2}\de(x_{21})|x_{31}|^{2(2j_1+n+1)}
|x_{32}|^{2(2j_2+n+1)}.
\end{equation}

Finally, one may observe that all terms that appear in the OPE are meromorphic
in $j_2$, $j_1$ with set of poles $\FS(j_2,j_1)$ given by 
\begin{equation}\label{sing}
2j_{21}^{\pm}=-(n+1)-(m+1)b^{-2}.
\end{equation}

\subsection{General remarks}
\subsubsection{} 
The leading term for $z_2\ra z_1$ appears in the first term of 
\rf{OPEfull}. It takes the following simple form:
\begin{equation}\label{leadas}\begin{aligned}
\Phi^{j_2} (x_2|z_2) \Phi^{j_1} & (x_1|z_1)= 
|z_2-z_1|^{-4b^2(j_1+1)(j_2+1)}\cdot \\
 & \cdot\Bigl( \de^2(x_2-x_1)
\Phi^{j_1+j_2+1}(x_1|z_1)
+\text{terms vanishing for $z_2\ra z_1$}\Bigr).
\end{aligned}
\end{equation}
\subsubsection{} It is worth noting that the state $\Psi_{21}\equiv
\Phi^{j_2}(x_2|z_2) \Phi^{j_1}(x_1|z_1)|0\ket$ will be in the domain of the
microscopic state $\bra -j_3-1|$ as defined in Section 2.3.3 
as long as one does not happen to hit one
of the poles \rf{polestot}. One then has
\begin{equation}
\bra -j_3-1|\Psi_{21}\ket=\lim_{z_3\ra\infty}|z|^{4\De_j}\bra 0|
\Phi^{j_3}(x_3|z_3)\Phi^{j_2}(x_2|z_2) \Phi^{j_1}(x_1|z_1)|0\ket,
\end{equation}
which ilustrates how the remarkable analyticity 
properties of the three point function make the distinction
between macroscopic and microscopic
states irrelevant for the issue of state-operator
correspondence.

\subsubsection{} The OPE gets singular for the values of $j_2$, $j_1$ 
given in \rf{sing}. This is due to the fact that the ``outgoing''
representation becomes degenerate for these values. In these cases there 
do not exist primary fields unless $j_2$, $j_1$ are further restricted
to satisfy the corresponding fusion rules.

\subsection{Interesting special cases}
\subsubsection{} For the special case $j_2=-1$ one thereby obtains the OPE 
of the operator that has appeared in \cite{KS} as the basic building block
for the construction of the space-time current- and Virasoro algebras.
The important point is that there does not appear a factor which
depends on $j_1$, as was conjectured in \cite{KS}. This is hereby
verified on the basis of the structure constants determined in \cite{T1}.
\subsubsection{} It is interesting to observe that the OPE simplifies
in the special case that $j_2$
and $j_1$ satisfy the so-called unitarity bound \cite{ES}\cite{EGP}
\begin{equation}
-1-\frac{1}{2b^2}<j_i<0,\qquad i=1,2
\end{equation}
and $k>3$. It only involves residue terms corresponding to
poles in $\CP_{21}^{\pm}$. Up to terms 
containing $\de^2(x_2-x_1)$ and derivatives thereof (``contact terms'')
one only finds terms containing
operators $\Phi^{-j_3-1}$, where $j_3$ also satisfies the
unitarity bound.
\subsubsection{} 
Another interesting case is that of integer level $k$.
The poles at $j_3\in\CQ_{21}^{\pm}$, $m>0$ 
degenerate into double poles for this
case, so that they do not lead to contributions in the OPE.

\newpage

\section{Factorization of the four point function} 

\subsection{Case with only macroscopic intermediate states} 
One may start with the case that 
\begin{equation}\label{funran2}\begin{aligned}
|\Re(j_{21}^{\pm})| & <\fr{1}{2}
\qquad\qquad j_{21}^+=j_2+j_1+1,\qquad j_{21}^{-}=j_2-j_1\\
|\Re(j_{43}^{\pm})| & <\fr{1}{2}
\qquad\qquad j_{43}^+=j_4+j_3+1,\qquad j_{43}^{-}=j_4-j_3.
\end{aligned}\end{equation}
In this case one may obtain the factorization 
e.g. in the s-channel one may use the
OPE \rf{OPE} for the pairs of operators $\Phi^{j_4}\Phi^{j_3}$ and
$\Phi^{j_2}\Phi^{j_1}$ and observe that only macroscopic intermediate
states are produced as a consequence of \rf{funran2}. 
 
The resulting decomposition of the four point function
can be written as (see Appendix B for some details)
\begin{equation}\label{factor1}
\bra\Phi^{j_4}\ldots\Phi^{j_1}\ket = \int_{\CC^+}dj\;\,
D_{43}(j)\;B(-j-1)\;D_{21}(j)
\; \CG_{j}(J|X|Z).
\end{equation}
The function $\CG_{j}(J|X|Z)$, $J=(j_4,\ldots,j_1)$, 
$X=(x_4,\ldots,x_1)$, $Z=(z_4,\ldots,z_1)$ that appears in \rf{factor1} 
will be called {\it non-chiral block}.
It takes the following form:
\begin{equation}\label{confbl1}\begin{aligned}
\CG_{j}(J|X|Z)= & \;
|z_{43}|^{2(\De_2+\De_1-\De_4-\De_3)}\;
|z_{42}|^{-4\De_2}\;|z_{41}|^{2(\De_3+\De_2-\De_4-\De_1)}
\cdot\\ 
\cdot & \;|z_{31}|^{2(\De_4-\De_1-\De_2-\De_3)}\;
|z|^{2\De_{21}(j)}\;  \CO_j(J|X|z)\overline{\CO_j(J|X|z)}
G_{j}(J|X),
\end{aligned}
\end{equation}
where the following objects have been introduced: There is first of all
the usual cross-ratio $z=\frac{z_{21}z_{43}}{z_{31}z_{42}}$.

The function $G_{j}(J|X)$ represents the summation over the
zero mode subspace $P_j$ of the representation $\CR_j$ and 
is given by the integral
\begin{equation}\label{zeroint}
G_{j}(J|X)=\int_{\BC}d^2 xd^2x'\;\, 
\frac{ C(j_4,j_3,j|x_4,x_3,x)C(j,j_2,j_1|x',x_2,x_1)}
{|x-x'|^{4j+4}}.
\end{equation}

The operators $\CO_j(J|X|z)$ finally are given as a formal power
series in $z$  
\begin{equation}\label{Oexp}
\CO_j(J|X|z)=\sum_{n=0}^{\infty}\;z^n \;\CD^{(n)}_j(J|X),
\end{equation}
where the $\CD^{(n)}_{x,j}(J|x)$ are differential operators 
containing derivatives w.r.t. $x_4,\ldots,x_1$ of finite order. These
operators are not known explicitly, but the construction given in Appendix
B will allow to obtain some important information.

\subsection{Chiral factorization}
\subsubsection{}
The integral \rf{zeroint} can be simplified by exploiting 
$SL(2,\BC)$-invariance 
\begin{equation}\label{projinv2}
G_{j}(J|X)= 
\;|x_{43}|^{2(j_4+j_3-j_2-j_1)}
\;|x_{42}|^{4j_2}\;|x_{41}|^{2(j_4+j_1-j_2-j_3)}\; 
|x_{31}|^{2(j_3+j_2+j_1-j_4)}\;
G_j(J|x),
\end{equation}
where $x=\frac{x_{21}x_{43}}{x_{31}x_{42}}$,
and the integral for $G_j(J|x)$ obtained by putting
$x_4=\infty$, $x_3=1$, $x_2=x$, $x_1=0$ can carried out by using
the integrals calculated in \cite{Do}, pp. 152:
\begin{equation}\label{Doint2}\begin{aligned}
G_j(J|x)=& \frac{\pi^2}{(2j+1)^2}\bigl|F_j(J|x)\bigr|^2 \\-
 & \frac{\pi^2}{\ga(2j+2)} 
\frac{\ga(1+j+j_4-j_3)\ga(1+j+j_3-j_4)}{\ga(j_1-j_2-j)\ga(j_2-j_1-j)}
\bigl|F_{-j-1}(J|x)\bigr|^2,
\end{aligned}\end{equation}
where $\ga(x)=\Ga(x)/\Ga(1-x)$ and 
\begin{equation}\label{zerodef}
F_j(J|x)\equiv x^{j_1+j_2-j}F(j_1-j_2-j,j_4-j_3-j;-2j;x).
\end{equation}
\subsubsection{}
The $\ga$-functions in front of the second term of \rf{Doint2} can be absorbed
by using \rf{refstruct}. 
It follows that 
the terms in \rf{factor1}
that originate from the second term in \rf{Doint2}  
are related to the terms from the first one by $j \ra -j-1$. This allows to
extend the integration over the half-axis $\CC^+$ in \rf{factor1} to an 
integration over the full axis $\CC=-\frac{1}{2}+i\BR$. The resulting 
expression shows a {\it holomorphically 
factorized} form
\begin{equation}\label{factor2}
\bra\Phi^{j_4}\ldots\Phi^{j_1}\ket = \int_{\CC}dj\;\; 
D_{43}(j)\;
\frac{1}{B(j)}\;D_{21}(j)
\;\bigl| \CF_{j}(J|X|Z)\bigr|^2,
\end{equation}
where the {\it chiral blocks} $\CF_{j_S}(J|X|Z)$ are of the form
\begin{equation}\label{confbl2}\begin{aligned}
\CF_{j}(J|X|Z)= & \;
z_{43}^{\De_2+\De_1-\De_4-\De_3}\;
z_{42}^{-2\De_2}\;z_{41}^{\De_3+\De_2-\De_4-\De_1}
\;z_{31}^{\De_4-\De_1-\De_2-\De_3}\;\cdot\\
 \cdot & \;x_{43}^{j_4+j_3-j_2-j_1}
\;x_{42}^{2j_2}\;x_{41}^{j_4+j_1-j_2-j_3}\;x_{31}^{j_1+j_2+j_3-j_4}F_j(J|x|z)\\
F_{j}(J|x|z) =& z^{\De_{21}(j)}\;  \CO_j(J|x|z) F_j(J|x).
\end{aligned}
\end{equation}

\subsection{Properties of conformal blocks}

\subsubsection{KZ-equations} The 4-point functions satisfy a system
of partial differential equations that is formally similar to the
Knizhnik-Zamolodchikov equations for the $SU(2)$-WZNW model (cf. Appendix C): 
\begin{equation}\label{KZ}
\pa_{z_r}F_j(J|X|Z)=\frac{1}{k-2}\sum_{r<s}\;\frac{\CD_{rs}}{z_r-z_s}\;\,
F_j(J|X|Z),
\end{equation}
where $r,s=1,\ldots,4$ and the differential operators $\CD_{rs}$ are given by
\begin{equation}
\CD_{rs}=-\CD^0_{j_r}\CD^0_{j_s}+\frac{1}{2}
\bigl(\CD^+_{j_r}\CD^-_{j_s}+\CD^-_{j_r}\CD^+_{j_s}\bigr)
\end{equation}
In the case of the four point function one may reduce it to
an equation for the cross-ratios 
$x$ and
$z$ which was first considered in the context of the
SU(2) WZNW model in \cite{FZ}:
\begin{equation}\label{kzred}
\pa_z F_j(J|X|Z) = 
\frac{1}{k-2}\left(\frac{P}{z}+\frac{Q}{z-1}\right)F_j(J|X|Z), \end{equation}
where ($\kappa \equiv j_1+j_2+j_3-j_4$)
\begin{equation}\begin{aligned}
P=&x^2(x-1)\pa_x^2-((\kappa-1)x^2    -2j_1x    +2j_2x(x-1))\pa_x
+2\kappa j_2x    -2j_1j_2 \\
Q=&-(1-x)^2x\pa_x^2+((\kappa-1)(1-x)^2-2j_3(1-x)+2j_2x(x-1))\pa_x
 \\
&+2\kappa j_2(1-x)-2j_3j_2 
\end{aligned}\end{equation}

\subsubsection{$X$- and $Z$-dependence of the chiral blocks}
The chiral blocks are expected to be multivalued analytic functions
of the variables $z_4,\ldots,z_1$ {\it and} $x_4,\ldots,x_1$ on 
$\BC^4\setminus \{z_r=z_s\}\ti\BC^4\setminus\{x_r=x_s\}$.
This will follow from the KZ-equations \rf{KZ} once the 
convergence of the formal power series that represent the
conformal blocks (cf. Appendix C) is established. 

Projective invariance allows to easily obtain the singularity structure
from the singularities of $F_j(J|x|z)$. These coincide with the 
singular points of \rf{kzred}, which are at $z=0,1,x,\infty$ and
$x=0,1,z,\infty$. The singular behavior near $z=0,1,\infty$ is of the
usual form as e.g. expressed by the last line in \rf{confbl2}.

Interesting is also the singular behavior near $x=0,1,z,\infty$. This
can also be determined from \rf{kzred}. One finds e.g.
\begin{equation}\begin{aligned} \label{xsing}
F_j(J|X|Z)\sim & \CO(1)+\CO(x_{21}^{j_1+j_2-j_3-j_4})\quad\text{near $x=0$},\\
F_j(J|X|Z)\sim & \CO(1)+\CO(x_{21}^{j_1+j_2+j_3+j_4+k})\quad\text{near $x=z$}.
\end{aligned}\end{equation}
The exponents at the singular points $x=0,1,z,\infty$ do not depend on $j$. 

\subsubsection{$J$- and $j$-dependence of the non-chiral blocks} 
An important consequence of the representation of the
chiral/non-chiral 
blocks in the form \rf{confbl1}\rf{confbl2} is that it allows to control the
$J$- and $j$-dependence of these objects. It follows from the 
construction (cf. Appendix B)
of the differential operators 
that appear in the expansion \rf{Oexp} of the operator $\CO_j(J|X|z)$ 
that $\CD^{(n)}_{x,j}(J|x)$ depends {\it polynomially} on $j_4,\ldots,j_1$
and {\it rationally} on $j$. $\CD^{(n)}_{x,j}(J|x)$ has only simple 
poles at the values $j=j_{r,s}^{\pm}$, $2j_{r,s}^{\pm}+1=\pm(r+b^{-2}s)$. 
Provided that the formal 
power series representation for $F_j(J|x|z)$ implied by \rf{confbl2},\rf{Oexp}
actually converges, one concludes that up to the poles at $j=j_{r,s}^{\pm}$
introduced by the descendant contributions one may read off the 
analyticity properties w.r.t. $J$ and $j$ from the lowest order
terms $G_{j}(J|X)$ or $F_j(J|X)$. In the case of $G_{j}(J|X)$
one finds from the expressions \rf{projinv2},\rf{Doint2} that the
set of poles in the $j$-dependence with positions depending 
on $J$ is just the union of the sets of poles of 
$C(j_4,j_3,j|x_4,x_3,x)$ and $C(j,j_2,j_1|x',x_2,x_1)$, which were discussed
in Section 3.1.3.

\subsection{Comments}

\subsubsection{}
The operators $\CO_j(J|X|z)$
are entirely determined by the
structure of the universal enveloping algebra $\CU(\hfg)$. In particular,
they do not depend on the choice of representation for the zero modes.

These operators may be seen as affecting the {\it deformation}
of the semiclassical chiral block $F_j(J|X)$ (cf. \cite{T2}) into 
the full conformal block $F_j(J|X|z)$.

Equivalently one may view it in connection to the KZ-equations as a kind
of {\it generalized wave operator} that maps solutions 
$z^{\De_{21}(j)}F_j(J|x)$ of the ``free wave equation''
$(k-2)z\pa_z F^{(0)}= P F^{(0)}$ into solutions $F_j(J|X|z)$ of \rf{kzred}. 

\subsubsection{}
In the case of the $H_3^+$-WZNW-model it 
is not a priori clear that the four point function posesses
a holomorphic factorization of the form \rf{factor2} since the 
representations of the current algebra that appear in the spectral 
decomposition \rf{absdec} do {\it not} show a holomorphic splitting as tensor 
product $\CV^j_L\ot \CV^j_R$, where $\CV^j_L$ ($\CV^j_R$) are 
irreducible representations of the current algebras generated by
the $J_n^a$ ($\bJ_n^a$) respectively. It is the 
representation of the zero modes $J_0^a$, $\bJ_0^a$ which does not
factorize as tensor product of seperate representations for $J_0^a$ and
$\bJ_0^a$ respectively. 

This is related to the fact that the integrand in \rf{factor2} for 
any single value does not contain all contributions of an intermediate
representation $\CR_j$. It is only the sum of the integrand taken 
at $j$ and $-j-1$ which represents the sum over contributions from 
$\CR_j$.
\subsubsection{}
The KZ-equations for the four point functions can not be reduced to
a system of ordinary differential equations as in the case of
the $SU(2)$-WZNW model studied in  \cite{FZ}. 

The problem
of constructing a complete (in an appropriate sense)
set of solutions to the KZ-equations
that can be identified with the chiral blocks 
is therefore rather nontrivial. This seems to be the main open problem
towards making the bootstrap approach to the $H_3^+$-WZNW model
mathematically rigorous. What can be shown at present (cf. Appendix C)
is that there exist unique
solutions in the sense of formal power series in $z$
which can be identified with chiral blocks. 

\subsubsection{}
The information on the singularities in the x-dependence
of four point functions 
is relevant for string theory on $ADS_3$ for the following reason:
If one tries to determine the OPE in the boundary CFT that
is supposed to describe the space-time physics of string theory on $ADS_3$,
one might consider e.g. the behavior of 
$V_{n_2}(x_2)V_{n_1}(x_1)$ for $x_2\ra x_1$. This should be encoded
in the four point function
\[
\CF(N|X)\equiv
\bra V_{n_2}(x_4)V_{n_1}(x_3)V_{n_2}(x_2)V_{n_1}(x_1)\ket,
\]
where $N=(n_2,n_1,n_2,n_1)$. The genus zero contribution to this 
four point function can be represented as 
\begin{equation}\label{intrep}\begin{aligned}
\CF(N|X)=\int d^2z \;\; & \bra \Phi^{j_2}(x_4|\infty)\Phi^{j_1}(x_3|1)
\Phi^{j_2}(x_2|z)\Phi^{j_1}(x_1|0)\ket_{H_3^+}^{}\cdot\\
\cdot & \bra \,\Xi_{n_2}(\infty)\,\Xi_{n_1}(1)\,\Xi_{n_2}(z)\,
\Xi_{n_1}(0)\ket_{\CN}^{},
\end{aligned}\end{equation}
where the $j_i$, $i=1,2$ are determined by the mass-shell condition in
\rf{vertop}, and $\bra\ldots\ket_{\CN}$ represents a correlation function
in the CFT describing $\CN$.

Equation \rf{xsing} now confirms the expectation \cite{KS} that singularities
of $\CF(N|X)$ for $x_2\ra x_1$ occur {\it only} from the part of the
region of integration in \rf{intrep} where $z\ra 0$, which 
corresponds to $\Phi^{j_2}(x_2|z_2)\Xi_{n_2}(z_2)$ approaching
 $\Phi^{j_1}(x_1|z_1)\Xi_{n_1}(z_1)$ on the {\it worldsheet}.

\subsection{Cases with microscopic intermediate states}

\subsubsection{}
If $j_4,\ldots,j_1$ are no longer restricted by \rf{funran2} 
then at least one of $\bra\Psi_{43}|\equiv \bra 0|\Phi^{j_4}\Phi^{j_3}$
and $|\Psi_{21}\ket\equiv \Phi^{j_2}\Phi^{j_1}|0\ket$will be 
microscopic.
 
The decomposition of the four point function as sum over  
non-chiral blocks in the case of generic $j_4,\ldots,j_1$ 
can be obtained by using the results of Section 3 for the
 OPE's $\Phi^{j_4}\Phi^{j_3}$ and
$\Phi^{j_2}\Phi^{j_1}$ in the generic case.
It follows from the remarks in Section 4.3.3 that 
the result will be just the same as obtained by directly
constructing the analytic continuation of the 
expression \rf{factor1} by the techniques of Section 3.

The resulting residue contributions are well-defined and non-singular
as long as the sets $\CS_{21}\equiv \CP^+_{21}\cup\CP^-_{21}
\cup \CQ^+_{21}\cup\CQ^-_{21}$ and $\CS_{43}$ are disjoint.
This condition is equivalent to conditions (1) and (2) of
Section 2.3.4 for the existence of the inner product 
$\bra\Psi_{43}|\Psi_{21}\ket$.

This shows that intermediate contributions of microscopic intermediate 
states are indeed perfectly well-defined in the generalization of
the bootstrap framework that was used in the present paper.

\subsubsection{}
The four point function develops poles for the set of $j_4,\ldots,j_1$
such that $\CS_{21}\cap\CS_{43}\neq\emptyset$. These cases are analogous
to the ``resonant'' amplitudes in Liouville theory (cf. e.g. \cite{FK}).
Representations like \rf{factor1} or \rf{factor2} for 
the residues at these poles can of course easily be worked out
from the results of the present paper. The integration over $j$ disappears
in these cases. It should be possible to
represent these residues more explicitly by free field methods.

\newpage

\section{Appendix A: Comparison of normalizations}

The aim of this appendix is to show that the normalization underlying
the expression \rf{Strconst} for the structure constants indeed coincides
with the one implied by the $\phi\ra\infty$-asymtotics \rf{opasym}.
This is furthermore compared to the normalization that was 
used in \cite{T1}.

\subsection{Comparison via leading order OPE}
\subsubsection{}
The leading order OPE can on the one hand be directly obtained from
the $\phi\ra\infty$-asymtotics \rf{opasym}: 
For the product of two operators $\Phi^{j_2}(x_2|z_2)$ and 
$\Phi^{j_1}(x_1|z_1)$ one finds that the $\phi\ra\infty$-asymtotics is given
by
\begin{equation}\label{asymprod}\begin{aligned}
\Phi^{j_2}(x_2|z_2) & \Phi^{j_1}(x_1|z_1)\sim  \\
 & \sim\de^2(x_2-x_1)
|z_2-z_1|^{-4b^2(j_1+1)(j_2+1)}\Bigl[\de^2(\ga-x_1):e^{2(-j_1-j_2-2)\Phi(z)}:
\Bigr],
\end{aligned}\end{equation}
where it was assumed that $j_i<-\frac{1}{2}$. The state created by
the operator on the right hand side will be non-normalizable 
provided that $j_1+j_2+1<-\frac{1}{2}$. In that case there is an unique
microscopic operator that has asymptotic behavior equal to the term
in square brackets that appears on 
the right hand side of \rf{asymprod}:
\begin{equation}
\Phi^{j_2}(x_2|z_2)\Phi^{j_1}(x_1|z_1)\sim
\de^2(x_2-x_1)
|z_2-z_1|^{-4b^2(j_1+1)(j_2+1)}\Phi^{j_1+j_2+1}(x_1|z_1).
\end{equation}
Terms that are subleading for $\phi\ra\infty$ lead to 
contributions in the OPE that are subleading for $z_2\ra z_1$.
\subsubsection{}
This should be compared with the leading order asymptotics that follows
from the structure constants \rf{Strconst} by using the methods of Sect. 3.
If one would keep $G_0$ and $\nu(b)$ in \rf{Strconst} arbitrary 
one would find instead of \rf{leadas} the expression
\begin{equation}\label{leadasgen}
\Phi^{j_2}(x_2|z_2)\Phi^{j_1}(x_1|z_1)\sim
2\pi^3\frac{G(-1)b^2}{G_0\nu(b)}\de^2(x_2-x_1)
|z_2-z_1|^{-4b^2(j_1+1)(j_2+1)}\Phi^{j_1+j_2+1}(x_1|z_1).
\end{equation}
This means that comparison of leading order asymptotics can only 
fix the product $G_0\nu(b)$.
\subsection{Comparison via two-point function}
The two-point function of the operators $\Theta^j(x|z)$ is  
recovered as the limit 
\begin{equation}\label{limit}
\lim_{\ep\ra 0}
\bra \Phi^{-\frac{1}{2}-i\rho'}(x_2|z_2)\Phi^{\ep}(x|z)
\Phi^{-\frac{1}{2}+i\rho}(x_1|z_1)\ket,
\end{equation}
which requires a little care due to the distributional nature of these
objects. The limit of course vanishes for $\rho\neq \rho'$ due to the factor 
$G^{-1}(2\ep+1)$. It will be assumed that $\rho,\rho'>0$ in the
following. There are
however two further terms that behave singular for 
$\rho-\rho'\simeq \ep\ra 0$: The factors $|x_3-x_1|^{2(i(\rho-\rho')-1-\ep)}$
and $G(\ep+i(\rho-\rho'))$. The singular behavior of the former can be
represented by $(i(\rho-\rho')-i\ep)^{-1}\pi\de^{(2)}(x_2-x_1)$, 
(cf. \rf{res}), whereas
the latter behaves as $(\ep+i(\rho-\rho'))^{-1}\Res_{x=0}G(x)$. 
The limit
\rf{limit} is therefore given by
\begin{equation}
|z_2-z_1|^{-4\De_{\rho}}\de^{(2)}(x_2-x_1)
\frac{G(-1)}{G_0}\frac{\Ga(1+b^2)}{\Ga(-b^2)}
\lim_{\ep\ra 0}\frac{2\ep}{\ep^2+(\rho-\rho')^2}.
\end{equation}
The last factor yields $2\pi\de(\rho-\rho')$, so that one would find the
two point function to be
\begin{equation}\label{twoptgen}
\bra \Phi^{-\frac{1}{2}-i\rho'}(x_2|z_2)
\Phi^{-\frac{1}{2}+i\rho}(x_1|z_1)\ket 
= -2\pi^2\frac{G(-1)}{G_0}\frac{\Ga(1+b^2)}{\Ga(-b^2)}\de(\rho-\rho')
\de^{(2)}(x_2-x_1)
\end{equation}
Requiring the prefactors in \rf{leadasgen} and \rf{twoptgen} to be unity
uniquely determines $G_0$ and $\nu(b)$ to be as given in Sect. 2.

\subsection{Normalization in \cite{T1}}
The normalization that was underlying the final result 
for the three point function \cite{T1} may be seen to be more natural
if one is interested in values of $j>-\frac{1}{2}$: In that case it is the
second term in \rf{opasym} that dominates the $\phi\ra\infty$-asymptotics,
so it is natural to require this term to have prefactor one. The operators
defined by this normalization will be denoted $\Theta^j(x|z)$.

The leading order OPE may then, as in the previous subsection, 
be determined from the $\phi\ra\infty$-asymptotics:
\begin{equation}\label{asymprodtheta}\begin{aligned}
\Theta^{j_2}(x_2|z_2) & \Theta^{j_1}(x_1|z_1)\sim  \\
 & \sim
|z_2-z_1|^{-4b^2(j_1+1)(j_2+1)}|\ga-x_2|^{4j_2}|\ga-x_1|^{4j_1}
:e^{2(j_1+j_2)\Phi(z)}:\\
 & \sim |z_2-z_1|^{-4b^2(j_1+1)(j_2+1)}\Bigl[ |\ga-x_1|^{4(j_1+j_2)}
:e^{2(j_1+j_2)\Phi(z)}:\Bigr]+\CO(x_2-x_1)\\
 & \sim |z_2-z_1|^{-4b^2(j_1+1)(j_2+1)}\Theta^{j_1+j_2}(x_1|z_1)+\CO(x_2-x_1)
\end{aligned}\end{equation}

This is to be compared to the result that follows from the structure 
constants of the operators $\Theta^j(x|z)$ by applying the method of 
Sect.3. These structure constants were given
in formula $(64)$ of \cite{T1}. In terms of the functions $G$ used in the
present paper one may rewrite the result as
\begin{equation}\label{Strtheta}
\tilde{D}( j_3 ,j_2,j_1)
=\frac{G(j_1+j_2+j_3+1)G(j_1+j_2-j_3)G(j_1+j_3-j_2)
G(j_2+j_3-j_1)}{(\nu'(b))^{-j_1-j_2-j_3-1} \;G_0' \;G(2j_1)G(2j_2)G(2j_3)}.
\end{equation}
The freedom that was left over by the analysis in \cite{T1} is parametrized
by $\nu'(b)$ and $G_0'$, which had not been determined there.

Before using the method of Section 3. to determine the
leading term in the OPE one needs to determine the 
precise form of the OPE in the case that $j_1$, $j_2$ are in the
range \rf{funran} where only macroscopic operators appear. To this aim 
one needs the two-point function following from \rf{Strtheta}. This is 
found to be 
\begin{equation}\label{twopttheta}
\bra \Theta^{-\frac{1}{2}-i\rho'}(x_2|z_2)
\Theta^{-\frac{1}{2}+i\rho}(x_1|z_1)\ket 
= \Bigl[
-2\pi^2b^{-4}\frac{G(-1)}{G_0'}\Bigr]
\frac{1}{4\rho^2}\de(\rho-\rho')
\de^{(2)}(x_2-x_1).
\end{equation}
Requiring the prefactor in square brackets to be unity determines $G_0'$.
It is important to note the factor $\frac{1}{4\rho^2}$ in \rf{twopttheta}:
Due to this factor one finds for the OPE of operators
$\Theta^{j_2}(x_2|z_2)
\Theta^{j_1}(x_1|z_1)$ in the case that $j_2$, $j_1$ are in \rf{funran}
a form that differs from \rf{OPE}:
\begin{equation}\label{OPETheta}\begin{aligned}
\Theta^{j_2}(x_2|z_2)
\Theta^{j_1}(x_1|z_1)
= -\frac{1}{2}
\int_{\CC}dj_3 \;  & (2j_3+1)^2   
 \;|z_2-z_1|^{\De_3-\De_2-\De_1}\tilde{D}(j_3,j_2,j_1)\cdot\\
& \cdot\int_{\BC} d^2x_3\; 
C(j_3,j_2,j_1|x_3,x_2,x_1)\;\Theta^{-j_3-1}(x_3|z_1)
\\
   & +\text{descendants},
\end{aligned}
\end{equation}
The measure $dj_3 (2j_3+1)^2$ is just the Plancherel-measure that appears in
the harmonic analysis on $H_3^+$ \cite{GGV},\cite{T2}.

By proceeding as in Sect.3 one now confirms that the 
leading term in the OPE is given by \rf{asymprodtheta}.
The operators $\Phi^j$ are related to the operators 
$\Theta^j$ by 
$\Phi^j(x|z)=B(j)\Theta^j(x|z)$.

\newpage

\section{Appendix B: Current algebra constraints on OPE and 
four point function}

For both the discussion of the OPE of two generic primary fields and the
description of the four-point function one needed to have some
information on the nature of the descendant contributions in OPE and
factorization. A more precise description of these contributions 
is made possible by means of the following useful technical tool:

\subsection{Dual basis}

\subsubsection{} 
To begin with, one needs to introduce some notation.
Given any basis $\{v_s; s\in\CS\}$ for the zero mode
subspace $P_j$ one may introduce  
an useful basis for $\CR_j$ as follows:
Let $\CT(n)$ be the set of all tuples 
\begin{equation}\label{tuple}
\mu=((r_1,\ldots,r_i,\ldots),(s_1,\ldots,s_j,\ldots),(t_1,\ldots,t_k,\ldots)) 
\end{equation}
such that 
\[ n=n(\mu)\equiv\sum_{i=1}^{\infty}ir_i+\sum_{j=1}^{\infty}js_j
+\sum_{k=1}^{\infty}kt_k \quad<\quad\infty \] 
Then the set of all
\begin{equation}\label{basis}
\FJ^{}_{\mu} \;v_s\equiv 
\prod_{i=1}^{\infty}(J_{-i}^{+})^{r_i}\prod_{j=1}^{\infty}(J_{-j}^{0})^{s_j}
\prod_{k=1}^{\infty}(J_{-k}^{-})^{t_k}\;\;v_{s}, \end{equation}
where $\mu\in\CT(n)$, $n\in\BN$ and $s\in\CS$ forms a basis for
$\CR_j$.
\subsubsection{}
The main result of this subsection is the construction of ``dual'' generators
$\FJ^{\mu}_{}\in\CU(\hfsl_2)$ that are specified by
\begin{equation}\label{dualdef}
\Pi_0\bigl(\FJ^{\mu}_{}\FJ^{}_{\nu}\bigr)=\de^{\mu}_{}{}_{\nu}^{} ,
\end{equation}
where $\Pi_0$ denotes projection onto the part that does not contain
terms of the form $(\ldots)J^a_n$ with $n>0$.
These ``dual'' generators will be of the form
\begin{equation}
\FJ^{\mu}_{}=\sum_{\nu\in \CT(n(\mu))}\;\; \FR^{\mu\nu}_{}(H,Q)\;\;\FJ_{\nu}^{\tau}
(J_0^{-\si(\mu,\nu)})^{\de(\mu,\nu)}_{},
\end{equation}
where the conjugation $\tau$ is defined by 
$(J_n^a)^{\tau}=J_{-n}^{-a}$ and $(AB)^{\tau}=B^{\tau}A^{\tau}$, 
$\si(\mu,\nu)=\sgn(m(\mu)-m(\nu))$, $\de(\mu,\nu)=|m(\mu)-m(\nu)|$,
$m(\mu)=\sum_{i=1}^{\infty}(r_i-t_i)$,
$H=J_0^0$ and $Q$ is the Casimir of the 
zero mode subalgebra, $Q=J^+_0J^-_0-J_0^0(J_0^0-1)$.

The crucial point about the matrix $ 
\FR^{\mu\nu}_{}(H,Q)$ is the fact that it depends {\it polynomially} on the
variable $H=J_0^0$ and rationally on the variable $Q$, with poles 
at $j=2j_{r,s}^{\pm}+1=\pm(r+b^{-2}s)$ when $Q=j(j+1)$.

This result gives a convenient representation of the identity on 
$\CV$ as
\begin{equation}\label{idrepr}
\Id=\int\limits_{\CC^+}
dj\;\;\sum_{\mu,\bar{\mu}}\;\;\int\limits_{\BC}d^2x\;\,
\FJ_{\mu}\BFJ_{\bar{\mu}}|j;x\ket\bra j;x|\FJ^{\mu}\BFJ^{\bar{\mu}}.
\end{equation}

\subsubsection{} 
The rest of this subsection will be devoted to the proof of this claim.
In order to construct the $\FJ^{\mu}_{}$ one may consider the generators
$\FJ_{\mu}^{}$ to be represented in Verma modules $\CV_j$
over the current algebra: The definition \rf{dualdef} is equivalent
to validity of
\begin{equation}\label{dualverma}
\FJ^{\mu}_{}\FJ^{}_{\nu}f_0=\de^{\mu}_{}{}_{\nu}^{}f_0 ,
\end{equation}
for any $f_0\in\CV_{j}$ such that $J^a_nf_0=0$ and generic $j$. The 
existence of ``dual'' generators $\FJ^{\mu}_{}$ for generic $j$
will then follow from 
the Kac-Kazhdan determinant formula \cite{KK}. What is not evident is the
claim on the polynomial dependence of the $\FJ^{\mu}_{}$ on the generator $H$. 

To control this dependence it is useful to choose as basis
for the Verma module the set 
\begin{equation}\{ e^j_{m,\nu}\equiv\FJ^{}_{\nu}e^j_{m-m(\nu)};\;\nu\in\CT;\; 
m=j,j-1,\ldots \},
\end{equation}
where the $e^j_m$ form a basis for the zero mode subspace $\CV_{j,0}=
\{ f_0\in\CV_j: \;J^a_nf_0=0 \}$ such that
\begin{equation}
J_0^+ e^j_m=-(m-j)e^j_{m+1}\qquad J_0^- e^j_m=-(m+j)e^j_{m-1}\qquad
H e^j_m=me^j_m.
\end{equation}
One may then introduce the bilinear form $(.,.)$ by
\begin{equation}
\bigl(\FJ^{}_{\mu}e^j_{m},\FJ^{}_{\nu}e^j_{n}\bigr)
=\bigl( e^j_{m}, \FJ^{\tau}_{\mu}\FJ^{}_{\nu}e^j_{n}\bigr)
\qquad (e^j_m,e^j_{n})=\de_{m,n}.
\end{equation}
This bilinear
form is {\it not} the usual Shapovalov form $\bra.,.\ket$
considered in \cite{KK}, but closely related to it:
\begin{equation}
\bra e^j_{m},e^j_{n}\ket
\bigl(\FJ^{}_{\mu}e^j_{m},\FJ^{}_{\nu}e^j_{n}\bigr)=
\bra\FJ^{}_{\mu}e^j_{m},\FJ^{}_{\nu}e^j_{n}
\ket
\bigl(e^j_{m},e^j_{n}\bigr).
\end{equation}
It follows that the determinant $D_{n}^j$ of the matrix 
$B_{\mu\nu}=(e^j_{m,\mu},e^j_{m,\nu})$, with $\mu,\nu\in\CT(n)$ is
proportional to the Kac-Kazhdan determinant:
\begin{equation}
D_{n}^{j}=N_{n}(k-2)^{R_3(n)}
\prod_{r,s\geq 1}\Bigl((j-j^+_{r,s})(j-j^-_{r,s})
\Bigr)^{P_3(n-rs)}, 
\end{equation} 
where $2j^{\pm}_{r,s}+1=\pm(r+b^{-2}s)$, $P_3(n)$ is the cardinality
of $\CT(n)$, $R_3(n)=\sum_{r,s\geq 1}P_3(n-rs)$. The prefactor
$N_n$ may be determined by 
considering the highest power in the central charge $k$, which is produced
by the product of the diagonal elements of the matrix $B_{\mu\nu}$.
The important point 
here is that it turns out to be m-independent for the chosen basis, which
implies $m$-independence of  $D_{n}^j$. 

\subsubsection{} The inverse of the matrix $B_{\mu\nu}$ therefore exists for 
$j\neq j_{r,s}^{\pm}$. The matrix elements of this inverse will be denoted
$B^{\mu\nu}$. It will be necessary to describe their $m$-dependence a
bit more precisely:

Let $s=-\si(\mu,\nu)$,
$I_{\mu\nu}=\{\de(\nu),\de(\nu)+s,\ldots,\de(\mu)-s\}$. 
$B^{\mu\nu}$ factorizes as
\begin{equation} \label{bsplit} 
 B^{\mu\nu}=\;
C^{\mu\nu}\prod_{k\in I_{\mu\nu}}(m-sj-k),
\end{equation}
where $C^{\mu\nu}=C^{\mu\nu}(m,j)$ is a polynomial in $m$.

This may be seen as follows:
It suffices to discuss the case $m(\nu)>m(\mu)$, the other case being 
completely analogous. One may equivalently fix a value of $k$
and ask in which elements $B^{\mu\nu}$ a factor $(m-j-k)$ can appear.

Now one may use the following simple fact: If for a matrix $M_{ij}$ the
upper right submatrix with lower left corner at $(k-1,k)$ carries a common
factor $a$, then the same will be true for the inverse matrix. This may
be verified by e.g. using the representation of the elements $M^{ij}$
of the inverse as $M^{ij}=(\det M)^{-1}\bar{M}^{ij}$ where $\bar{M}^{ij}$
is the determinant of the matrix obtained from $M$ 
by erasing the $i^{th}$ row and $j^{th}$ column. The contributions to these
determinants may be associated to zig-zag paths that go in unit steps 
from the right to the left in the matrix $\bar{M}^{ij}$
and which never hit the same vertical position
twice. The matrix element $M^{ij}$ will carry a factor $a$ if there
exists no path that avoids the upper right submatrix in $\bar{M}^{ij}$
with lower left corner at $(k-1,k)$. This is in fact the case if
$i<k\leq j$: The lower right submatrix in $\bar{M}^{ij}$ with
upper left corner $(k,k)$ then has less columns than rows.
 
It therefore suffices to verify that the factor $(m-j-k)$ appears in 
any matrix element $B_{\mu\nu}$
with $m(\mu)<k\leq m(\nu)$. But this follows from
the definition by observing that
\[
\FJ^{\tau}_{\mu}\FJ_{\nu}e^j_{m-m(\nu)}= 
R_{\mu\nu}(H,Q) (J_0^+)_{}^{m(\nu)-m(\mu)}
e^j_{m-m(\nu)},
\]
where $R_{\mu\nu}(H,Q)$ is some polynomial in the indicated variables.

\subsubsection{}
The factorization \rf{bsplit} now allows to rewrite the 
identity on the Verma module $\CV_j$ as follows:
\begin{equation}\begin{aligned}
\Id=&\sum_{m=-\infty}^j \sum_{\mu,\nu}
\FJ_{\mu}^{}e^j_{m-m(\mu)}
B^{\mu\nu}(m,j)\bigl(\FJ_{\nu}^{}e^j_{m-m(\nu)}\bigr)^{\tau}\\
=& \sum_{m=-\infty}^j \sum_{\mu,\nu}\FJ_{\mu}^{}e^j_{m-m(\mu)}
C^{\mu\nu}(m,j)\bigl((J_0^{-\si(\mu,\nu)})_{}^{\de(\mu,\nu)}e^j_{m-m(\mu)}
\bigr)^{\tau}\FJ_{\nu}^{\tau},
\end{aligned}
\end{equation}
where the summation over $\mu,\nu$ is restricted by $m-m(\mu)\leq j$
and $m-m(\nu)\leq j$. The summations over 
$\mu,\nu$ and $m$ may be exchanged, giving a sum over $m$ that is 
restricted by the conditions just mentioned. The summation over $m$
may then be shifted by $m(\mu)$, where restricting the new sum by
$m-m(\nu)+m(\mu)\leq j$ is superfluous since automatically realized
due to $J^+ e^j_j=0$. 
The result may be written as
\begin{equation}
 \Id=\sum_{\mu}\sum_{m=-\infty}^j \FJ_{\mu}^{}e^j_m\bigl(e^j_m\bigr)^{\tau}
\FJ^{\mu}_{},
\end{equation}
which is equivalent to the desired equation \rf{dualverma} with dual generators
$\FJ^{\mu}$ being given by
\begin{equation}
\FJ^{\mu}_{}\equiv \sum_{\nu}\;\;C^{\mu\nu}\bigl(H+m(\mu),Q\bigr)
\;\,\FJ_{\nu}^{\tau}
\;\,(J_0^{\si(\mu,\nu)})_{}^{\de(\mu,\nu)}
\end{equation}

\subsection{Descendant operators}
\subsubsection{}
The descendants of the states $|j;x\ket$ with fixed $j$ represent elements
of the dual $\CR_j{}^t$ of $\CR_j$.
The aim will be to associate an operator $\Phi^j(\gv|z)$ to 
each distribution $\gv\in\CR_j{}^t$. This can be done by defining
recursively
\begin{equation}\label{descdef}
\Phi^j(\de_x|z)\equiv\Phi^j(x|z) \qquad\qquad
\begin{aligned}
\Phi^j\bigl(J_{-n}^a\gv|z\bigr)=& \frac{1}{(n-1)!}:\pa_z^{n-1}J^a(z)
\Phi^j(\gv|z):,\\
\Phi^j\bigl(\bJ_{-n}^a\gv|z\bigr)=& \frac{1}{(n-1)!}:\pa_{\bz}^{n-1}\bJ^a(z)
\Phi^j(\gv|z):,
\end{aligned}\end{equation}
where $\de_x\in P_j{}^t$ represents the delta-functional
$\de_x(f)=f(x)$ for any $f\in P_j\simeq \CS(\BC)$.
Normal ordering is defined here by
$:J^a(z)\CO(z):= J_<^a(z)\CO(z)+\CO(z)J_>^a(z)$
where 
\[ J_<^a(z)=\sum_{n<0}J^a_nz^{-n-1},\qquad 
J_>^a(z)=\sum_{n\geq 0}J^a_nz^{-n-1}. \]
State-operator correspondence \rf{stopprim} generalizes to 
\begin{equation}\label{stopsec}
\lim_{z\ra 0}\Phi^{j}\bigl(\FJ_{\mu}\bar{\FJ}_{\bar{\mu}}\de_x|z\bigr)
|0\ket = 
\FJ_{\mu}\bar{\FJ}_{\bar{\mu}}|j;x\ket \qquad\text{or}\qquad
e^{zL_{-1}}\FJ_{\mu}\bar{\FJ}_{\bar{\mu}}|j;x\ket=
\Phi^{j}\bigl(\FJ_{\mu}\bar{\FJ}_{\bar{\mu}}\de_x|z)|0\ket.
\end{equation}
The transformation properties of descendants under the current algebra 
are given by the commutation relations
\begin{equation}\label{sectrsf}
[J^a_>(w),\Phi^j(\gv|z)]=\Phi^j\bigl(J^a_>(w-z)\gv|z\bigr)\qquad
[J^a_<(w),\Phi^j(\gv|z)]=-\Phi^j\bigl(J^a_>(w-z)\gv|z\bigr).
\end{equation}
One should note that $J^a_0\de_x=\CD_j^a\de_x$, so that \rf{sectrsf}
includes the transformation law \rf{prim2} as a special case.

\subsubsection{}
The conformal Ward identities are now a consequence of the definition
\rf{descdef}, commutation relations \rf{sectrsf} and 
invariance of the vacuum $|0\ket$: One may express any correlation 
function of descendant operators $\bra \Phi^{j_N}(\gv_N|z_N)\ldots 
\Phi^{j_1}(\gv_1|z_1)\ket$ in terms of $\bra \Phi^{j_N}(x_N|z_N)\ldots 
\Phi^{j_1}(x_1|z_1)\ket$ by 
expressing any chosen $\Phi^{j_i}(\gv_i|z_i)$ in terms of 
$\Phi^{j_i}(x_i|z_i)$ with the help of \rf{descdef},
moving the $J_<^a$, $\bJ_<^a$, (resp. $J_>^a$, $\bJ_>^a$) to the left 
(resp. right) vaccuum by using \rf{sectrsf}, and then repeating 
the process until only primary fields are left.  

In the case of the two point function of two 
descendant operators one needs to supplement these definitions 
by requiring it to vanish when the levels of the 
descendants of the two operators are unequal.
\begin{equation}\label{twoptdesc}\begin{aligned}
\bra \;\Phi^{-j_2-1} & \bigl(\FJ_{\mu}\bar{\FJ}_{\bar{\mu}}\de_{x_2}|z_2\bigr)
 \Phi^{j_1}\bigl(\FJ_{\nu}\bar{\FJ}_{\bar{\nu}}\de_{x_1}|z_1\bigr)\;\ket=\\
& = \;|z_{21}|^{-4\De_{j_1}}z_{21}^{-2n(\mu)}z_{21}^{-2n(\bar{\mu})}\;
\de_{n(\mu),n(\nu)}\de_{n(\bar{\nu}),n(\bar{\mu})}\;
\bra j_2;x_2|\FJ_{\mu}^{\si}\FJ_{\bar{\mu}}^{\si}\FJ_{\nu}^{}
\FJ_{\bar{\nu}}^{}|j_1;x_1\ket,
\end{aligned}\end{equation}
where the conjugation $\si$ acts as $(J_n^a)^{\si}=J_{-n}^{a}$.

It follows from \rf{twoptdesc} that out-states are created via
\begin{equation}
\bra j;x|(\FJ_{\mu}\bar{\FJ}_{\bar{\mu}})^{\si}=
\lim_{z\ra\infty}
|z|^{4\De_{j_1}}z^{2n(\mu)}z^{2n(\bar{\mu})}\bra 0|
\Phi^{j}\bigl(\FJ_{\mu}\bar{\FJ}_{\bar{\mu}}\de_{x}|z\bigr)
\end{equation}

\subsection{Descendant contributions in OPE and factorization}
\subsubsection{}
The resolution of the identity \rf{idrepr}
immediately gives an expansion for the ``in''-state
$\Phi^{j_2}(x_2|z_2)|j_1;x_1\ket$ with coefficients given 
by the matrix elements $\bra j;x|\FJ^{\mu}\bar{\FJ}^{\bar{\mu}}
\Phi^{j_2}(x_2|z_2)|j_1;x_1\ket$. 
This is brought into correspondence
with \rf{OPE} by using the $j_3\ra-j_3-1$ symmetry
of the integrand (cf. 3.1.1). One gets an expansion of the form 
\begin{equation}\label{inexp}\begin{aligned}
\Phi^{j_2}(x_2|z_2) &  |j_1;x_1\ket=\\
& \int\limits_{\CC^+}
dj_3  \;\sum_{\mu,\bar{\mu}}\;
D(j_3,j_2,j_1)z^{\De_{21}+n(\mu)}\bz^{\De_{21}+n(\bar{\mu})}\cdot \\
& \qquad\quad \cdot\int\limits_{\BC}d^2x_3\;\,\bigl(\CD_{x_2}^{j_2}(\FJ^{\mu})
\CD_{\bx_2}^{j_2}(\bar{\FJ}^{\bar{\mu}})
C(j_3,j_2,j_1|x_3,x_2,x_1)\bigr)|-j_3-1,x_3\ket,
\end{aligned}
\end{equation}
where the differential operators $\CD_{x_2}^{j_2}(\FJ^{\mu})$ are obtained by 
replacing in the expression for the dual generators $\FJ^{\mu}(j_3)$ all 
generators $J_n^a$ that appear by the differential operators 
$\CD^a_{x_2}(j_2)$.

\subsubsection{}
The full operator product expansion is now easily 
obtained by applying the translation operator 
$e^{z_1L_{-1}}$ to \rf{inexp} and using \rf{stopsec}:
\begin{equation}\label{OPEdesc}\begin{aligned}
{}&\Phi^{j_2}(x_2| z_2) \Phi^{j_1}(x_1|z_1)
= \\
& =
\int\limits_{-\frac{1}{2}+i\BR^+}
dj_3   \;\;D(j_3,j_2,j_1)\sum_{\mu,\bar{\mu}}
 \;|z_{21}|^{2\De_{21}(j_3)}
z_{21}^{n(\mu)}\bz_{21}^{n(\bar{\mu})}
\cdot\\[-1ex]
& \qquad\qquad\cdot\int_{\BC} d^2x_3\; \;
\Bigl(\CD_{x_2}^{j_2}(\FJ^{\mu})\CD_{\bx_2}^{j_2}(\FJ^{\bar{\mu}})\;
C(j_3,j_2,j_1|x_3,x_2,x_1)\Bigl)\;\Phi^{j_3}
\bigl(\FJ_{\mu}\bar{\FJ}_{\bar{\mu}}\de_{x_3}|z_1\bigr).
\end{aligned}
\end{equation}
\subsubsection{}
It is worth noting that the polynomial dependence of the $\FJ^{\mu}$ on the
zero mode generators $J^a_0$ is of course crucial for getting 
differential operators of finite order in \rf{OPEdesc}.

It also follows that the descendant contributions do not introduce any
$j_2,j_1$-dependent pole of the OPE-coefficients
in addition to the poles discussed in Section 3.1. 

\subsubsection{Four point function}
Projective invariance determines the $z$-dependence of the four point 
function to be of the form
\begin{equation}\begin{aligned}
\bra\Phi^{j_4}(x_4|z_4)\ldots \Phi^{j_1}(x_1|z_1)\ket = & \;
|z_{43}|^{2(\De_2+\De_1-\De_4-\De_3)}\;
|z_{42}|^{-4\De_2}\;|z_{41}|^{2(\De_3+\De_2-\De_4-\De_1)}
\cdot\\ 
\cdot & \;|z_{31}|^{2(\De_4-\De_1-\De_2-\De_3)}\;\;
\bra j_4;x_4|\Phi^{j_3}(x_3|1)
\Phi^{j_2}(x_2|z)|j_1;x_1\ket,
\end{aligned}
\end{equation}
where $z$ is the usual cross ratio defined below \rf{confbl1}.
In order to get information on the $F_j(J|X|z)$ is therefore suffices to
consider the case $z_4=\infty$, $z_3=1$, $z_2=z$, $z_1=0$.

A representation of the four point function as sum over intermediate states
can be obtained e.g. by 
expanding $\Phi^{j_2}(x_2|z)|j_1;x_1\ket$ as in \rf{inexp},
using \rf{idrepr} to get a similar 
expansion for the ``out''-state $\bra j_4;x_4|\Phi^{j_3}(x_3|1)$
and finally using the inner products \rf{normstates}. One obtains a 
representation of the for point function as sum over products of
three point functions which takes the form
\begin{equation}\label{fourint}\begin{aligned}
\bra \Phi^{j_4}\ldots \Phi^{j_1}\ket= &
\int\limits_{-\frac{1}{2}+i\BR^+}dj\;\;\sum_{\mu,\bar{\mu}}\;\; 
z^{\De_{21}+n(\mu)}\bz^{\De_{21}+n(\bar{\mu})} D_{43}(j)B(-j-1)D_{21}(j)
\cdot \\
& \qquad\qquad\qquad\cdot\tilde{\CD}_{x_3}^{j_3}(\FJ_{\mu})
\tilde{\CD}_{\bx_3}^{j_3}(\BFJ_{\bar{\mu}})
\CD_{x_2}^{j_2}(\FJ^{\mu})\CD_{\bx_2}^{j_2}(\BFJ^{\bar{\mu}})
\;G_j(J|X),
\end{aligned}
\end{equation}
where the function $G_j(J|X)$ is given by the integral \rf{zeroint}
and the differential operators $\tilde{\CD}_x^j(\FJ_{\mu})$) are 
obtained by replacing any generator $J_n^a$
that appears in $\FJ_{\mu}$ by the differential operator $-\CD^a_{j,x}$.

The differential operators $\CD^{(n)}_j(J|X)$
that appear in \rf{Oexp} are the finally defined 
by summing the differential operators  appearing in \rf{fourint}
over $\mu$ with $n(\mu)=n$.

\section{Appendix C:Knizhnik-Zamolodchikov equations}

The derivation of the KZ-equations for the $H_3^+$-WZNW model is very 
similar to the case of WZNW-models for compact groups, see
e.g. \cite{FR} for a derivation that can be easily adapted to
the formalism used in Appendix B.
It is based on the result that the operators satisfy
the following operator differential equation in the sense of formal 
power series:
\begin{equation}\label{KZop}
\frac{d}{dz}\Phi^j(\de_x|z)=\frac{1}{2(k-2)}\sum_{a,b=+,0,-}\eta_{ab}
\Phi^j\bigl(J^a_{-1}J^{b}_0\de_x|z\bigr),
\end{equation}
where $\eta_{00}=-2$, $\eta_{+-}=1=\eta_{-+}$ are the only nonvanishing
matrix elements of $\eta_{ab}$.

When inserting \rf{KZop} into correlation functions and using 
the conformal Ward identities one finds the KZ-equations \rf{KZ}. 

\subsection{Solutions to KZ}

Solutions that can be identified with s-channel 
conformal blocks must be of the form :
\[ F_{j}(J|x|z)=
z^{\De_{21}(j)}\;\sum_{n=0}^{\infty}\; z^n\;
F^{(n)}_{j}(J|x)
\]
The KZ-equation \rf{kzred} then implies the following set of 
recursion relations for the coefficients $F^{(n)}_{j}(J|x)$:
\begin{equation}
\label{KZrec1} \bigl(\De_{21}+n+(k-2)^{-1}P\bigr)
F^{(n)}_{j}(J|x)=-\frac{1}{k-2}Q
\sum_{m=0}^{n-1}F^{(n)}_{j}(J|x) \end{equation}
For $n=0$ one has the eigenvalue equation 
$(P+(k-2)\De_{21})F^{(0)}_{j}(J|x)=0$. Two linearly independent solutions
are $F_j(J|x)$ and $F_{-j-1}(J|x)$, cf. \rf{zerodef}.

The equations with $n>0$
are inhomogeneous differential equations that uniquely determine 
$F^{(n)}_{j}(J|x)$ in terms of $F^{(m)}_{j}(J|x)$, $m<n$
up to solutions of the homogeneous equation
$(\De_{21}+n+(k-2)^{-1}P)f=0$. However, the corresponding
ambiguity can be fixed by observing that on the one hand
solutions which can be identified with conformal blocks have to be of the
form 
\begin{equation}\label{KZcoeff}
 F^{(n)}_{j}(J|x)=x^{j_1+j_2-j-\ka(n)}\sum_{m=0}^{\infty}
\;x^m\;F^{(n,m)}_{j}(J),  \end{equation}
where $\ka(n)$ is integer as follows from \rf{KZrec1}.
But on the other hand one may observe that
the exponents $\la_n$ of solutions to the homogeneous equation 
$(\De_{21}+n+(k-2)^{-1}P)f=0$ at the singular point $x=0$ are 
determined by the equation 
\[  \bigl(\la_n-(j_1+j_2+\fr{1}{2})\bigr)^2=(j
+\fr{1}{2})^2-(k-2)n,  \]
so if one writes $\la_n=j_1+j_2-j-l(n)$ one finds that $(2j+1)l(n)=(k-2)n$.
Unless $2j+1$ is a rational multiple of $k-2$ one finds that $l(n)$ 
is not an integer so that adding a solution to the homogeneous equation
to a solution of \rf{KZrec1} would prevent the identification of the
solution with a conformal block.

Furthermore one has to check that there always exists a solution of the 
required form (\ref{KZcoeff}). 
First observe that  
In order to satisfy (\ref{KZrec1}) the coefficients $F_{j}^{(n,m)}(J)$
will have to satisfy recursion relations of the following form:
\[
A_m F_{j}^{(m,n)}+B_m F_{j}^{(n,m-1)}(J)=
G_{j}^{(n,m-1)}(J) \]
where the coefficients are given by
\[\begin{array}{rcl}
A_m &=& \left(\tau_n+m-\left(j_1+j_2+\frac{1}{2}\right)\right)^2-  
\left(j_{21}+\frac{1}{2}\right)^2+(k-2)n \\
B_m &=& (\tau_n+m-\kappa)(\tau_n+m-2j_2)\end{array};\qquad \tau_n\equiv 
j_1+j_2-j_{21}-n 
\]
and the $G_{j}^{(n,m-1)}(J)$ are defined by
\[
x^{\tau_n}\sum_{m=0}^{\infty}\;\;x^m\;\;G_{j}^{(n,m-1)}(J)\equiv 
Q\sum_{i=0}^{n-1}F^{(s)}_{j_{21},n}(x)  
\]
The recursion relations are solvable unless there is 
some $m$ for which $A_m=0$. But 
$A_m=(m-n)^2-(2j_{21}+1)(m-n)+
(k-2)n$, which will not vanish for any integer $m$ unless 
$2j+1$ is a rational multiple of $k-2$.

\newcommand{\CMP}[3]{{\it Comm. Math. Phys. }{\bf #1} (#2) #3}
\newcommand{\LMP}[3]{{\it Lett. Math. Phys. }{\bf #1} (#2) #3}
\newcommand{\IMP}[3]{{\it Int. J. Mod. Phys. }{\bf A#1} (#2) #3}
\newcommand{\NP}[3]{{\it Nucl. Phys. }{\bf B#1} (#2) #3}
\newcommand{\PL}[3]{{\it Phys. Lett. }{\bf B#1} (#2) #3}
\newcommand{\MPL}[3]{{\it Mod. Phys. Lett. }{\bf A#1} (#2) #3}
\newcommand{\PRL}[3]{{\it Phys. Rev. Lett. }{\bf #1} (#2) #3}
\newcommand{\AP}[3]{{\it Ann. Phys. (N.Y.) }{\bf #1} (#2) #3}
\newcommand{\LMJ}[3]{{\it Leningrad Math. J. }{\bf #1} (#2) #3}
\newcommand{\FAA}[3]{{\it Funct. Anal. Appl. }{\bf #1} (#2) #3}
\newcommand{\PTPS}[3]{{\it Progr. Theor. Phys. Suppl. }{\bf #1} (#2) #3}
\newcommand{\LMN}[3]{{\it Lecture Notes in Mathematics }{\bf #1} (#2) #2}

\end{document}